\documentclass[fleqn,useAMS,usenatbib]{mnras}

\usepackage[T1]{fontenc}
\usepackage{ae,aecompl}
\usepackage{xspace}
\usepackage{graphicx}  
\usepackage{footnote}
\makesavenoteenv{tabular}
\makesavenoteenv{table}
\makesavenoteenv{sidewaystable}
\usepackage[fleqn]{amsmath}
\usepackage{amsfonts}
\usepackage{amssymb}
\usepackage[]{xcolor}

\usepackage{srcltx}

\usepackage{euclid}

\allowdisplaybreaks[4]
\bibpunct{(}{)}{;}{a}{}{,}

\def\figrelpath{./}
\def\bibpath{refs}

\newcommand{\satellitename}[1]{\textit{#1}}
\newcommand{\telescopename}[1]{\text{#1}}
\newcommand{\softwarename}[1]{\textsc{#1}}

\newcommand{\vect}[1]{\boldsymbol{#1}}

\newcommand{\ev}[1]{\left\langle #1 \right\rangle}

\newcommand{\ft}[1]{\hat{#1}}

\DeclareMathOperator{\sinc}{sinc}

\newcommand{\dd}{\mathrm{d}}
\newcommand{\ii}{\mathrm{i}}
\newcommand{\Ee}{\mathrm{e}}

\newcommand{\PS}[3]{P^{#1}_{#2}\left(#3\right)}
\newcommand{\harmPS}[3]{C^{#1}_{#2}\left(#3\right)}

\newcommand{\kpc}{\ensuremath{\mathrm{kpc}}}
\newcommand{\Mpc}{\ensuremath{\mathrm{Mpc}}}

\newcommand{\arcsect}{\ensuremath{\mathrm{arcsec}}}
\newcommand{\arcmint}{\ensuremath{\mathrm{arcmin}}}
\newcommand{\degt}{\ensuremath{\mathrm{deg}}}

\newcommand{\clight}{\ensuremath{\mathrm{c}}}
\newcommand{\GNewton}{\ensuremath{\mathrm{G}}}
\newcommand{\HubbleConstant}{H_{0}}

\newcommand{\OmegaMatter}{\Omega_{\mathrm{m}}}
\newcommand{\OmegaBaryon}{\Omega_{\mathrm{b}}}

\newcommand{\deltaMatter}{\delta_{\text{m}}}
\newcommand{\corrMatter}{\xi_{\text{m}}}
\newcommand{\corrMatterperp}{\xi_{\text{m}}^{\perp}}

\newcommand{\PSMatter}[2]{\PS{#1}{\text{m}}{#2}}

\newcommand{\gammat}{\gamma_{\text{\textsc{t}}}}
\newcommand{\gammax}{\gamma_{\!\times\!}}

\newcommand{\vvartheta}{\vect{\vartheta}}
\newcommand{\vtheta}{\vect{\theta}}
\newcommand{\vbeta}{\vect{\beta}}

\newcommand{\vell}{\vect{\ell}}

\newcommand{\zD}{z_{\mathrm{d}}}
\newcommand{\zS}{z_{\mathrm{s}}}

\newcommand{\chiD}{\chi_{\mathrm{d}}}
\newcommand{\chiS}{\chi_{\mathrm{s}}}

\newcommand{\fK}{f_{K}}
\newcommand{\fD}{f_{\mathrm{d}}}
\newcommand{\fS}{f_{\mathrm{s}}}
\newcommand{\fDS}{f_{\mathrm{ds}}}

\newcommand{\qDS}{q_{\mathrm{ds}}}

\newcommand{\euclid}{\satellitename{Euclid}}
\newcommand{\wfirst}{\satellitename{WFIRST}}
\newcommand{\lsst}{\telescopename{LSST}}

\newcommand{\nside}{N_\text{side}}

\newcommand{\ramses}{\softwarename{RAMSES}}
\newcommand{\rayramses}{\softwarename{Ray-RAMSES}}

\newcommand{\lptic}{\softwarename{2LPTic}}
\newcommand{\pinocchio}{\softwarename{Pinocchio}}
\newcommand{\icecola}{\softwarename{ICE-COLA}}
\newcommand{\peakpatch}{\softwarename{PeakPatch}}
\newcommand{\mapsim}{\softwarename{MapSim}}
\newcommand{\fabbian}{\softwarename{LenS$^2$HAT}}
\newcommand{\hilbert}{\softwarename{Hilbert}}
\newcommand{\camb}{\softwarename{CAMB}}
\newcommand{\halofit}{\softwarename{Halofit}}

\newcommand{\lenstools}{\softwarename{Lenstools}}
\newcommand{\glamer}{\softwarename{Glamer}}
\newcommand{\mice}{\softwarename{MICE}}

\newcommand{\healpix}{\softwarename{HEALPix}}
\newcommand{\healpy}{\softwarename{healpy}}

\begin{document}

\title[The Accuracy of WL Simulations]{The Accuracy of Weak Lensing Simulations}
\author[S. Hilbert, et al.]{Stefan Hilbert$^{1,2}$\thanks{\href{mailto:stefan.hilbert@tum.de}{\texttt{stefan.hilbert@tum.de}}},
Alexandre Barreira$^{3}$,
Giulio Fabbian$^{4,5}$,
Pablo Fosalba$^{6, 7}$, \newauthor
Carlo Giocoli$^{8,9,10}$,
Sownak Bose$^{11}$,
Matteo Calabrese$^{12}$,
Carmelita Carbone $^{13}$, \newauthor
Christopher T. Davies $^{14}$,
Baojiu Li$^{14}$,
Claudio Llinares$^{15}$,
Pierluigi Monaco$^{16,17,18,19}$
\\$^{1}$ Exzellenzcluster Universe, Boltzmannstr. 2, D-85748 Garching, Germany,
\\$^{2}$ Ludwig-Maximilians-Universit{\"a}t, Universit{\"a}ts-Sternwarte, Scheinerstr. 1, D-81679 M{\"u}nchen, Germany,
\\$^{3}$ Max-Planck-Institut f{\"u}r Astrophysik, Karl-Schwarzschild-Str. 1, D-85748 Garching, Germany,
\\$^{4}$ Department of Physics \& Astronomy, University of Sussex, Brighton BN1 9QH, UK,
\\$^{5}$ Institut d'Astrophysique Spatiale, CNRS (UMR 8617), Univ. Paris-Sud, Universit\'{e} Paris-Saclay, B\^{a}t. 121, 91405 Orsay, France,
\\$^{6}$ Institute of Space Sciences (ICE, CSIC), Campus UAB, Carrer de Can Magrans, s/n, 08193 Barcelona, Spain
\\$^{7}$ Institut d'Estudis Espacials de Catalunya (IEEC), 08034 Barcelona, Spain
\\$^{8}$ Dipartimento di Fisica e Astronomia, Alma Mater Studiorum Universit\`{a} di Bologna, via Gobetti 93/2, I-40129 Bologna, Italy,
\\$^{9}$ INAF, Osservatorio di Astrofisica e Scienza dello Spazio di Bologna,  via Gobetti 93/3, I-40129 Bologna, Italy,
\\$^{10}$ INFN, Sezione di Bologna, viale Berti Pichat 6/2, I-40127 Bologna, Italy,
\\$^{11}$ Center for Astrophysics | Harvard \& Smithsonian, 60 Garden Street, Cambridge, MA 02138, USA
\\$^{12}$ Astronomical Observatory of the Autonomous Region of the Aosta Valley (OAVdA), Loc. Lignan 39, I-11020, Nus (AO), Italy
\\$^{13}$ INAF - Institute of Space Astrophysics and Cosmic Physics (IASF) Via Corti 12, I-20133 Milano (MI), ITALY
\\$^{14}$ Institute for Computational Cosmology, Department of Physics, Durham University, South Road, Durham DH1 3LE, UK
\\$^{15}$Institute of Cosmology and Gravitation, University of Portsmouth, Dennis Sciama Building, Portsmouth PO1 3FX, United Kingdom
\\$^{16}$ Universit\`a di Trieste, Dipartimento di Fisica, sezione di Astronomia, via Tiepolo 11, I-34143 Trieste, Italy,
\\$^{17}$ INAF-Osservatorio Astronomico di Trieste, via Tiepolo 11, I-34143 Trieste, Italy,
\\$^{18}$ INFN, Sezione di Trieste, via Valerio 2, 34127 Trieste, Italy
\\$^{19}$ IFPU - Institute for the Fundamental Physics of the Universe, Via Beirut 2, 34014, Trieste, Italy
}


\date{\today}

\pubyear{20??}

\label{firstpage}
\pagerange{\pageref{firstpage}--\pageref{lastpage}}
\maketitle

\begin{abstract}
We investigate the accuracy of weak lensing simulations by comparing the results of five independently developed lensing simulation codes run on the same input $N$-body simulation. Our comparison focuses on the lensing convergence maps produced by the codes, and in particular on the corresponding PDFs, power spectra and peak counts. We find that the convergence power spectra of the lensing codes agree to $\lesssim 2\%$ out to scales $\ell \approx 4000$. For lensing peak counts, the agreement is better than $5\%$ for peaks with signal-to-noise $\lesssim 6$. We also discuss the systematic errors due to the Born approximation, line-of-sight discretization, particle noise and smoothing. The lensing codes tested deal in markedly different ways with these effects, but they nonetheless display a satisfactory level of agreement. Our results thus suggest that systematic errors due to the operation of existing lensing codes should be small. Moreover their impact on the convergence power spectra for a lensing simulation can be predicted given its numerical details, which may then serve as a validation test.
\end{abstract}


\section{Introduction}
\label{sec:introduction}

The images of distant galaxies are weakly sheared due to the differential deflection of their light by the gravity of the intervening cosmic  large-scale structure. This gravitational lensing effect is commonly referred to as weak lensing or cosmic shear \citep[see, e.g.,][for reviews]{2001PhR...340..291B,2015RPPh...78h6901K,2018ARA&A..56..393M}, and it carries information about both the space-time geometry and the large-scale matter distribution of the Universe.
Cosmic shear observations therefore prove extremely useful in tests of cosmological models and constraints on cosmological parameters, as the analyses of the Kilo Degree Survey \citep[KiDS,][]{2017MNRAS.465.1454H, 2018arXiv181206076H}, the Dark Energy Survey \citep[DES,][]{2018PhRvD..98d3526A, 2018PhRvD..98d3528T}, and the Subaru Hyper Suprime-Cam (HSC) survey \citep[][]{2019PASJ...71...43H, 2019arXiv190606041H} have recently demonstrated.

Upcoming wide-field imaging surveys such as the \euclid{} satellite \citep{2011arXiv1110.3193L} \footnote{\href{http://www.euclid-ec.org}{\texttt{http://www.euclid-ec.org}}}, the Large Synoptic Survey Telescope \citep{2012arXiv1211.0310L} \footnote{\href{http://www.lsst.org}{\texttt{http://www.lsst.org}}} (\lsst{}), or the \satellitename{Wide Field Infrared Survey Telescope} \citep{2013arXiv1305.5422S} \footnote{\href{https://wfirst.gsfc.nasa.gov}{\texttt{https://wfirst.gsfc.nasa.gov}}} (\wfirst{}) will allow us to measure the cosmic shear signal over a wide range of angular scales and redshifts with very small statistical uncertainties \citep[e.g.][]{2011arXiv1110.3193L}. To fully exploit this unprecedented statistical power, we require not only a thorough understanding of the systematic errors inherent in the measurements themselves, but also very accurate theoretical predictions for cosmic structure formation and its associated cosmic shear signal. Numerical simulations are an extremely valuable and widespread tool in gravitational lensing analysis. They can be used to compute predictions for gravitational lensing observables accurately in the nonlinear/small-scale regime of structure formation (which can be used to calibrate faster semi-analytical methods to compute gravitational lensing predictions), as well as to build synthetic mock lensing data to be used in tests of different methods to measure the lensing shear signal from observations. Lensing simulations are, however, also subject to statistical and systematic errors themselves, and these must be well understood in order to appropriately apply and interpret their output. 

Various methods for simulating gravitational lensing observations have been developed over the past few decades. They differ in the intended application, which naturally leads to differences in, for instance, the way the deflector mass is distributed and modeled, or in the method to compute the gravitational light deflection and distortion caused by the deflecting mass. For example, stars as deflectors in microlensing studies have been modeled as point masses \citep[e.g.][]{1986ApJ...301..503P,1990ApJ...352..407W,2004ApJ...605...58K}. Analytic extended mass profiles have been used to represent the lensing mass of galaxies and clusters in strong lensing simulations \citep[e.g.][]{1987ApJ...321..658B, 1988ApJ...324L..37G,1998A&A...330..399B,2001ApJ...563....9M,2002ApJ...573...51O, 2012MNRAS.421.3343G,2015MNRAS.447.3189X,2018MNRAS.475.5424D}. Other simulations of the strong lensing signal caused by galaxies and clusters employ \lq{}non-parametric\rq{} mass distributions extracted from $N$-body and hydrodynamical simulations, either alone or in combination with analytic mass profiles \citep[e.g.][]{1994A&A...287....1B,2000MNRAS.314..338M, 2017MNRAS.472.3177M,2005ApJ...633..768H, 2011MNRAS.418...54H, 2005A&A...442..405P,2009MNRAS.398.1298P, 2007MNRAS.382..121H, 2008MNRAS.386.1845H,2009MNRAS.398.1235X}.

There is also a variety of methods to simulate the weak lensing signal caused by the large-scale distribution of matter in the Universe, which is the signal we focus on in this paper. Some simulations adopt analytical prescriptions \citep[e.g.][]{2011PhRvD..84f3004K, giocoli17} or realizations of lognormal distributed fields \citep[e.g.][]{2016MNRAS.459.3693X} to describe the large-scale mass distribution. A more common approach has been to use the matter distribution generated by $N$-body or hydrodynamical simulations of cosmic structure formation. Among these, many methods employ a multiple-plane \citep[e.g.][]{1998ApJ...494...29W,2000ApJ...530..547J,2001MNRAS.327..169H,2009AaA...499...31H,2012MNRAS.426.1262H,2014MNRAS.445.1954P, 2016A&C....17...73P} or multiple-sphere algorithm \citep[e.g.][]{fosalba08, dasbode2008,2013MNRAS.435..115B,fabbian2018, 2019A&A...626A..72G}, in which the continuous mass distribution is projected onto a set of discrete two-dimensional mass distributions that act as deflectors along the line-of-sight. There are also methods that bypass the line-of-sight discretization and make use of the three-dimensional mass distribution of the $N$-body simulation outputs \citep[e.g.][]{1999MNRAS.308..180C,2003ApJ...592..699V,carbone2008,2011MNRAS.415..881L, 2011MNRAS.414.2235K,2016JCAP...05..001B,2019MNRAS.483.2671B}. 

The different weak-lensing simulation methods have different advantages and disadvantages in terms of speed and memory requirements, but importantly, the type and number of approximations made in them can also have an impact on their accuracy. Even for numerical codes that adopt overall the same lensing simulation method, there can still be differences in the final result, as different implementations may handle different approximations differently. Published work on weak lensing simulations usually contains tests of the correctness and accuracy of the numerical algorithms, but what is exactly tested and the way it is reported can vary widely, which makes comparisons across the literature hard. The importance to establish benchmark tests of the accuracy of these methods is therefore hard to overstate, specially given the ever increasing precision of the observational data \citep[][]{2017arXiv170609359K}.

Specifically in this paper, we investigate the current level of accuracy of weak lensing simulations by comparing the results of different lensing simulation codes ran on the same output of an $N$-body simulation of cosmic structure formation. We compare five codes: \hilbert{} \citep{2007MNRAS.382..121H, 2009AaA...499...31H} and \mapsim{} \citep{2015MNRAS.452.2757G}, which are post-processing multiple-plane lensing codes; \mice{} \citep{fosalba08, fosalba15} and \fabbian{} \citep{fabbian2013, calabrese2015}, which are post-processing multiple-sphere codes; and \rayramses{} \citep{2016JCAP...05..001B}, which runs on-the-fly with the simulation and uses the three-dimensional distribution directly. We focus the comparison on the lensing convergence maps, their associated power spectra and PDFs, as well as lensing peaks counts. We also comment on systematic errors such as those associated with the Born approximation, particle noise, smoothing, and the line-of-sight discretization for multiple-plane/sphere methods. Based on these results, we outline and discuss a procedure to help validate the output of weak lensing simulations and quantify their accuracy.

The code comparison we present in this paper adds to a large  body of work on the validation of numerical $N$-body codes and algorithms to extract cosmological information from their output. Some of the codes/methods already subject to similar testing include: gravity-only $N$-body algorithms \citep{2016JCAP...04..047S}, including non-standard gravity \citep{2015MNRAS.454.4208W}, galaxy formation codes \citep{2012MNRAS.423.1726S}, methods to identify halos/subhalos \citep{2011MNRAS.415.2293K, 2012MNRAS.423.1200O}, galaxies \citep{2013MNRAS.428.2039K}, voids \citep{2008MNRAS.387..933C} and tidal debris \citep{2013MNRAS.433.1537E} from the output of $N$-body simulations, codes to construct halo merger trees \citep{2013MNRAS.436..150S} and fast generation of halo catalogues  \citep{2015MNRAS.452..686C}, including comparing the covariances of their two- \citep{2019MNRAS.482.1786L, 2019MNRAS.485.2806B} and three-point statistics \citep{2019MNRAS.482.4883C}. The importance of validation analyses such as these is crucial to identify and mitigate sources of systematic errors in our theoretical predictions. This serves as the main motivation for the weak-lensing code comparison analysis carried out here.

This paper is organized as follows. In Section~\ref{sec:theory}, we review the main theoretical aspects of weak gravitational lensing and how accurately can one expect numerical simulation methods to operate. In Section~\ref{sec:methods}, we describe the setup of our code comparison analysis, including the underlying $N$-body simulation of cosmic structure and the operation of the lensing simulation codes that participate in the comparison. Our main comparison results are presented and discussed in Sec.~\ref{sec:results}, where we also quantify systematic errors at play in lensing simulations. We summarize and conclude in Section~\ref{sec:conclusions}.

\section{Theory}
\label{sec:theory}

In this section, we display some of the main equations needed to understand weak-lensing observables, with emphasis on the calculation of two-point statistics, including how it can be affected by numerical resolution issues in lensing simulations (such as discretization of the line-of-sight and particle shot-noise).

\subsection{Weak gravitational lensing}
\label{sec:theory:wl}

While traveling from source to observer, light will be deflected by the gravity of intervening matter structures \citep[see, e.g.,][ for reviews]{2001PhR...340..291B,2015RPPh...78h6901K, 2018ARA&A..56..393M}. In a weakly perturbed Friedmann-Lema{\^i}tre-Robertson-Walker universe, the angular (unobserved) position $\vbeta$ of a source at comoving line-of-sight distance $\chiS$ and redshift $\zS = z(\chiS)$ and with (observed) image position $\vtheta$ on the sky is given to very high accuracy by:
\begin{equation}
\label{eq:lens_equation}
 \vbeta (\vtheta, \zS) = \vtheta - \frac{2}{\clight^2} \int_0^{\chiS} \diff{\chiD} \frac{\fDS}{\fD \fS}
 \nabla_{\vbeta} \Phi\bigl(\vbeta (\vtheta, \chiD), \chiD, \zD \bigr).
\end{equation} 
Here, $\clight$ denotes the speed of light, $\fDS = \fK(\chiS - \chiD)$, $\fD = \fK(\chiD)$ and $\fS = \fK(\chiS)$, where $\fK(\chi)$ denotes the comoving angular diameter distance for comoving  line-of-sight distance $\chi$, and $\zD = z(\chiD)$ the redshift corresponding to comoving line-of-sight distance $\chiD$.
Furthermore, $\nabla_{\vbeta}$ denotes the gradient w.r.t.~the angular position $\vbeta$, and $\Phi\bigl(\vbeta, \chiD, \zD \bigr)$ denotes the Newtonian gravitational potential at position $(\vbeta, \chiD)$ and redsfhift $\zD$.
On a flat sky, the Jacobian\footnote{For simplicity, we discuss these equations for a flat sky with $\vbeta$ and $\vtheta$ as two-dimensional Cartesian coordinate vectors. For a spherical sky, the partial derivatives w.r.t.~the angular positions have to be replaced by covariant derivatives on the sphere \citep{2013MNRAS.435..115B}.}
\begin{equation}
\label{eq:lens_jacobian}
 \frac{\partial \vbeta}{\partial \vtheta} = 
 \begin{pmatrix}
  1 - \kappa - \gamma_1 & -\gamma_2 - \omega \\
  -\gamma_2 + \omega & 1 - \kappa + \gamma_1 \\
 \end{pmatrix}
\end{equation} 
defines the lensing convergence $\kappa$, lensing shear $\gamma = \gamma_1 + \ii \gamma_2$, and the lensing rotation $\omega$.
From Eq.~\eqref{eq:lens_equation} it follows that
\begin{multline}
\label{eq:lens_jacobian_equation}
 \frac{\partial \beta_i (\vtheta, \zS)}{\partial \theta_j} =
 \delta_{ij} - \frac{2}{\clight^2} \int_0^{\chiS} \diff{\chiD}
 \frac{\fDS}{\fD \fS}
\\\times 
\frac{\partial^2 \Phi\bigl(\vbeta (\vtheta, \chiD), \chiD, \zD \bigr)}{\partial \beta_i \partial \beta_k}
 \frac{\partial \beta_k\bigl(\vtheta, \chiD \bigr)}{\partial \theta_j},
\end{multline}
where $\delta_{ij}$ is the Kronecker delta symbol. Replacing $\vbeta$ on the r.h.s. by the unperturbed position $\vtheta$ in Eq~\eqref{eq:lens_jacobian_equation} yields the first-order (in $\Phi$) approximation:
\begin{equation}
\label{eq:lens_jacobian_equation_1st_order}
 \frac{\partial \beta_i (\vtheta, \zS)}{\partial \theta_j} = 
\delta_{ij}
 - \frac{2}{\clight^2} \int_0^{\chiS} \diff{\chiD}
 \frac{\fDS}{\fD \fS}
 \frac{\partial^2 \Phi\bigl(\vtheta, \chiD, \zD \bigr)}{\partial \theta_i \partial \theta_j},
\end{equation}
which is sometimes called the Born approximation.\footnote{
The term \lq{}Born approximation\rq{} is not used uniformly throughout the literature, and sometimes instead refers to an equation obtained by just replacing $\partial_{\beta_i}\partial_{\beta_k}\Phi(\vbeta, \chiD, \zD)$ by  $\partial_{\theta_i}\partial_{\theta_k} \Phi(\vtheta, \chiD, \zD)$ in Eq.~\eqref{eq:lens_jacobian_equation}, but keeping the factor $\partial_{\theta_j} \beta_k$.
}
Note that the rotation $\omega$ vanishes in this approximation.

If one makes use of the Poisson equation for the gravitational potential $\Phi$ and also neglects boundary terms at the observer and source, one obtains the following approximation for the convergence:
\begin{equation}
\label{eq:convergence_1st_order}
 \kappa (\vtheta, \zS) = 
\int_0^{\chiS} \diff{\chiD}
\,\qDS\,
\deltaMatter\bigl(\vtheta, \chiD, \zD \bigr)
\end{equation}
with the lensing efficiency factor
\begin{equation}
\label{eq:lensing_efficiency_factor}
 \qDS = 
\frac{3 \HubbleConstant^2 \OmegaMatter}{2 \clight^2}
(1 + \zD) \fD\frac{\fDS}{\fS}
,
\end{equation}
where $\deltaMatter\bigl(\vtheta, \chiD, \zD \bigr)$ is the matter density contrast.

Assuming statistical isotropy, the two-point correlation $\xi_\kappa(\vartheta, \zS)$ of the convergence $\kappa$ for sources at angular separation $\vartheta$ and redshift $\zS$ can then be written as:
\begin{equation}
\label{eq:xi_kappa_1st_order}
\begin{split}
\xi_\kappa(|\vvartheta|, \zS)
 &= 
\ev{\kappa(\vtheta, \zS) \kappa(\vtheta + \vvartheta, \zS)} 
\\&=
\int_0^{\chiS} \diff{\chiD}
\,\qDS\,
\int_0^{\chiS} \diff{\chiD'}
\,\qDS'\,
\\&\quad\times
\ev{
\deltaMatter\bigl(\vtheta, \chiD, \zD \bigr)
\deltaMatter\bigl(\vtheta + \vvartheta, \chiD', \zD' \bigr)
}
.
\end{split}
\end{equation}
Here, $\ev{\ldots}$ denotes the expectation for a given (statistically homogeneous and isotropic) ensemble of universes.

Assuming $\deltaMatter$ is slowly evolving with redshift and matter correlations are short-ranged compared to $\chiS$, one can apply a Limber-type approximation to obtain:
\begin{equation}
\label{eq:xi_kappa_1st_order_Limber}
\xi_\kappa(\vartheta, \zS)
 = 
\int_0^{\chiS} \diff{\chiD}
\,\qDS^2\,
\corrMatterperp\bigl(\vartheta \chiD, \zD \bigr)
,
\end{equation}
where
\begin{equation}
\corrMatterperp\bigl(R, z \bigr)
=
\int_{-\infty}^{\infty} \diff L\, \corrMatter\bigl(\sqrt{R^2 + L^2}, z \bigr)
,
\end{equation}
is the line-of-sight projection of the three-dimensional two-point correlation function $\corrMatter(r, z)$ of the matter density contrast at comoving separation $r$ at redshift $z$. The corresponding approximation to the spherical harmonic power spectrum $\harmPS{}{\kappa}{\ell}$ of the convergence as a function of harmonic wave number $\ell$ reads:
\begin{equation}
\label{eq:ps_kappa_1st_order_Limber}
\harmPS{}{\kappa}{\ell}
 = 
\int_0^{\chiS} \diff{\chiD}
\,\qDS^2\,
\PSMatter{}{\ell / \chiD, \zD}
,
\end{equation}
where $\PSMatter{}{k, \zD}$ denotes the three-dimensional matter power spectrum for wave number $k$ at redshift $\zD$.

In cosmic-shear surveys, one aims to estimate the shear from the observed shapes of galaxy images. 
Given a direction $\vvartheta$ on the sky, one may define a tangential and a cross shear component as
\begin{subequations}
\begin{align}
\label{eq:gamma_t}
\gammat(\vtheta, \zS, \vvartheta) &= \Re\left[-\Ee^{-2 \ii \varphi(\vvartheta)} \gamma(\vtheta, \zS) \right],
\\
\label{eq:gamma_x}
\gammax(\vtheta, \zS, \vvartheta) &= \Im\left[-\Ee^{-2 \ii \varphi(\vvartheta)} \gamma(\vtheta, \zS) \right],
\end{align}
\end{subequations}
where $\varphi(\vvartheta)$ is the position angle of $\vvartheta$. Using these, one may define the shear correlation functions
\begin{equation}
\label{eq:xi_gamma_pm}
\begin{split}
\xi_\pm(|\vvartheta|, \zS)
&= \ev{\gammat(\vtheta, \zS, \vvartheta) \gammat(\vtheta + \vvartheta, \zS, \vvartheta)} 
\\&\quad  
\pm \ev{\gammax(\vtheta, \zS, \vvartheta) \gammax(\vtheta + \vvartheta, \zS, \vvartheta)}.
\end{split}
\end{equation}
Within the first-order approximation, we have that
\begin{subequations}
\begin{align}
\xi_+(\vartheta, \zS) &= \xi_\kappa(\vartheta, \zS)
\\
\begin{split}
\xi_-(\vartheta, \zS) &= \xi_\kappa(\vartheta, \zS) 
\\&\quad 
+ 
 \int_0^{\vartheta}\dd\vartheta'\,
\left(\frac{4\vartheta'}{\vartheta^2}-\frac{12\vartheta^{\prime 3}}{\vartheta^4}\right)
\xi_{\kappa} (\vartheta', \zS)
.
\end{split}
\end{align}
\end{subequations}

The shear $\gamma$ transforms like a spin-2 field and thus can be decomposed into rotationally-invariant E and B-modes \citep{stebbins96,kks97,zs97}. 
When employing the flat-sky, first-order (Born), and Limber approximations, the shear E- $\harmPS{\text{(EE)}}{\gamma}{\ell}$ and B-mode $\harmPS{\text{(BB)}}{\gamma}{\ell}$ power spectra obey the following relations:
\begin{subequations}
\label{eq:ps_gamma_1st_order_Limber}
\begin{align}
\label{eq:ps_gamma_E_1st_order_Limber}
\harmPS{\text{(EE)}}{\gamma}{\ell}
&=
  \harmPS{}{\kappa}{\ell},\\
\label{eq:ps_gamma_B_1st_order_Limber}
\harmPS{\text{(BB)}}{\gamma}{\ell} &= \harmPS{}{\omega}{\ell} = 0.
\end{align}
\end{subequations}
where $\harmPS{}{\omega}{\ell}$ is the lensing rotation power spectrum.

Corrections to the flat-sky approximation for the convergence and shear power spectra become relevant only on very large scales with $\ell < 10^2$ \citep[e.g.][]{2000PhRvD..62d3007H, 2005PhRvD..72b3516C, 2013MNRAS.435..115B, 2017MNRAS.472.2126K}. For example, the relation between the convergence and shear power spectrum is modified by a factor $[(\ell + 2)(\ell - 1)]/[\ell(\ell+ 1)]$ on a spherical sky, which differs by $<1\%$ from unity for $\ell \geq 15$. Corrections to the Limber approximation also only become relevant on very large scales. For $\ell \gtrsim 10^2$, such corrections to the convergence and shear power spectra are expected to be below $1\%$ \citep[][]{2015MNRAS.448..364A, 2017MNRAS.472.2126K}.

Second- and higher-order contributions in the gravitational potential $\Phi$ to the convergence and shear yield small corrections to their power spectra. These are expected to be at least two orders of magnitude below the leading order terms for $10^2 \lesssim \ell \lesssim 10^4$ \citep[see, e.g.,][]{2002ApJ...574...19C, 2003PhRvD..68h3002H, 2006JCAP...03..007S, 2009AaA...499...31H, 2010A&A...523A..28K,2017PhRvD..95l3503P}. Note, however, that beyond-Born corrections may be more significant for other weak-lensing quantities. In particular, the galaxy-galaxy lensing shear profiles \citep[e.g.][]{2008PhRvD..78l3517Z,2009AaA...499...31H,2018A&A...613A..15S,2018JCAP...06..008G}, higher-order moments of the convergence field \citep[][]{2017PhRvD..95l3503P} and lensing bispectrum \citep{pratten2016} may require one to go beyond first order and undeflected light rays \citep{petri17}.

Higher-order terms also cause a non-vanishing rotation $\omega$, and create a small B-mode component in the shear field that is likely below the detection threshold even for upcoming galaxy lensing surveys. The amplitude of these corrections depends however on the distance to the lensing sources (the farther the sources the higher the {\it number of deflections}), which has been motivating studies of its importance for future analysis of the lensing of the cosmic microwave background \citep{pratten2016,lewis2016,marozzi2016, fabbian2018}.

\subsection{Weak lensing power spectra in simulations}
\label{sec:theory:wl_sims}

The convergence and shear power spectra measured from simulated weak-lensing observations may differ from the pure theory predictions~\eqref{eq:ps_kappa_1st_order_Limber} and \eqref{eq:ps_gamma_1st_order_Limber} for a number of reasons. First, there is sample variance because every simulation set covers only a finite number of realizations of a given area fraction and depth. The impact of sample variance can however be estimated analytically, by resampling techniques, or by generating many realizations of the simulated lensing maps.

Second, there is the issue of the intrinsic accuracy of Eqs.~\eqref{eq:ps_kappa_1st_order_Limber} and \eqref{eq:ps_gamma_1st_order_Limber}, which assume a flat sky and use the Born and the Limber approximation. As we noted in the previous subsection, corrections from going beyond these approximations on the convergence and shear power spectra are however expected to be well below $1\%$ for $10^2 \lesssim \ell \lesssim 10^4$, which is sufficiently small even for surveys like \euclid, \lsst, or \wfirst. Hence, taking the validity of these approximations as established thus allows one to actually use them in internal self-consistency tests: codes ran in and out of these approximations should return spectra that differ comfortably by less than $1\%$ on the relevant scales.

Third, the methods to measure the convergence and shear from the simulations may introduce biases. In this work, we do not consider  possible issues due to, e.g., differences between shear $\gamma$ and reduced shear $g = \gamma/(1-\kappa)$ (which is more closely related to observed galaxy ellipticities) or noisy and biased image shape measurements. Instead we assume that we have bias- and noise-free measurements of the convergence (except in Sec.~\ref{sec:results:peak_counts}, when we also add Gaussian random noise to the simulated convergence maps before we count lensing peaks).

Fourth, numerical approximations (smoothing, line-of-sight projections, particle discreteness, etc.) employed in the lensing simulations may impact the simulated convergence and shear power spectra in various ways. However, with sufficient knowledge of the numerical details of the simulations, one may be able to account for some of these effects in a modified theory prediction for the convergence and shear power spectra. One may then compare these modified predictions with the measured power spectra as part of a validation procedure for the lensing simulations \citep[see, e.g., Sec.~3 of][]{fosalba08}.

As an example, consider a lensing simulation based on an $N$-body simulation of cosmic structure formation. In that case, the matter power spectrum $\PSMatter{\text{(sim)}}{k,z}$ going into the lensing simulation can be described by a continuous component and a shot noise component due to sampling of the density field by a finite number of particles/mass elements in the simulation:
\begin{equation}
\label{eq:P_matter_sim}
\PSMatter{\text{(sim)}}{k,z}  = \PSMatter{\text{(cont)}}{k,z} + \PSMatter{\text{(sn)}}{k,z}.
\end{equation}
The continuous component $\PSMatter{\text{(cont)}}{k,z}$ should closely resemble the theoretical power spectrum $\PSMatter{}{k,z}$ at least on the scales represented well by the $N$-body simulation, but deviates more strongly on very large and very small scales due to the finite simulation box size and resolution. The shot-noise term $\PSMatter{\text{(sn)}}{k,z}$ is usually sufficiently well described by white noise with an amplitude given by the inverse particle density, $\PSMatter{\text{(sn)}}{k,z} = V N_{\text{p}}^{-1}$, where $V$ denotes the simulation volume, and $N_{\text{p}}$ denotes the simulation particle number (assumed to have the same mass).

As a first step to adjust the prediction for the convergence power spectra from the simulation, one may replace the theoretical power spectrum with the power spectrum measured from the $N$-body simulation:
\begin{equation}
\label{eq:limberPk}
\harmPS{\text{(sim)}}{\kappa}{\ell}
 = 
\int_0^{\chiS} \diff{\chiD}
\,\qDS^2\,
\PSMatter{\text{(sim)}}{\ell / \chiD, \zD}
.
\end{equation}
Further, if the matter distribution of the simulation is only available at a finite number of snapshots at redshifts, $\zD^{(i)}$, $i=1,\ldots$, one can write
\begin{equation}
\harmPS{\text{(sim)}}{\kappa}{\ell}
 = 
\sum_{i}
\int_{\chiD^{(i,\text{lo})}}^{\chiD^{(i,\text{hi})}} \diff{\chiD}
\,\qDS^2\,
\PSMatter{\text{(sim)}}{\ell / \chiD, \zD^{(i)}}
,
\end{equation}
where $\chiD^{(i,\text{lo})}$ and $\chiD^{(i,\text{hi})}$ denote the lower and upper boundary of the slice of the observer's backward lightcone filled by the matter distribution from snapshot $i$ at redshift $\zD^{(i)}$.
If the lensing simulation employs lens planes at distances $\chiD^{(i)}$ with one plane per snapshot and with the matter projected onto lens planes with parallel projection, then we can write
\begin{equation}
\label{eq:limberPk_planes}
\begin{split}
\harmPS{\text{(sim)}}{\kappa}{\ell}
 &= 
\sum_{i}
\left(\chiD^{(i,\text{hi})} - \chiD^{(i,\text{lo})}\right)
\\&\quad\times
{\qDS^{(i)}}^2\,
\PSMatter{\text{(sim)}}{\ell / \chiD^{(i)}, \zD^{(i)}}
,
\end{split}
\end{equation}
where ${\qDS^{(i)}}$ is the lensing efficiency at $\zD^{(i)}$. Finally, if all of the binning, interpolation and smoothing schemes employed by the lensing simulations in their various steps can be effectively described by some convolution of the three-dimensional matter distribution with a window function $W$ with a fixed comoving smoothing scale, then the prediction for the simulated convergence power spectrum can be written as:
\begin{equation}
\label{eq:limberPk_planes_and_smoothing}
\begin{split}
\harmPS{\text{(sim)}}{\kappa}{\ell}
 &= 
\sum_{i}
\left(\chiD^{(i,\text{hi})} - \chiD^{(i,\text{lo})}\right)
\\&\quad\times
{\qDS^{(i)}}^2\,
\ft{W}(\ell / \chiD^{(i)})
\PSMatter{\text{(sim)}}{\ell / \chiD^{(i)}, \zD^{(i)}}
,
\end{split}
\end{equation}
where $\ft{W}(k)$ denotes the Fourier transform of filter $W$ (see Sec.~\ref{sec:results:systematic_errors:particle_noise_and_smoothing} for a specific application).

\section{Methods}
\label{sec:methods}

In this section, we describe the numerical methods that participate in this comparison project. This includes the lensing simulation codes themselves (\hilbert{}, \fabbian{}, \mapsim{}, \mice{}, and \rayramses{}), as well as the $N$-body simulation that provides the common realization of large-scale structure on which they perform their calculations.

\subsection{$N$-body simulation and lightcone geometry}
\label{sec:simulations:N_body_simulations}

\begin{figure*}
\centerline{\includegraphics[width=\linewidth]{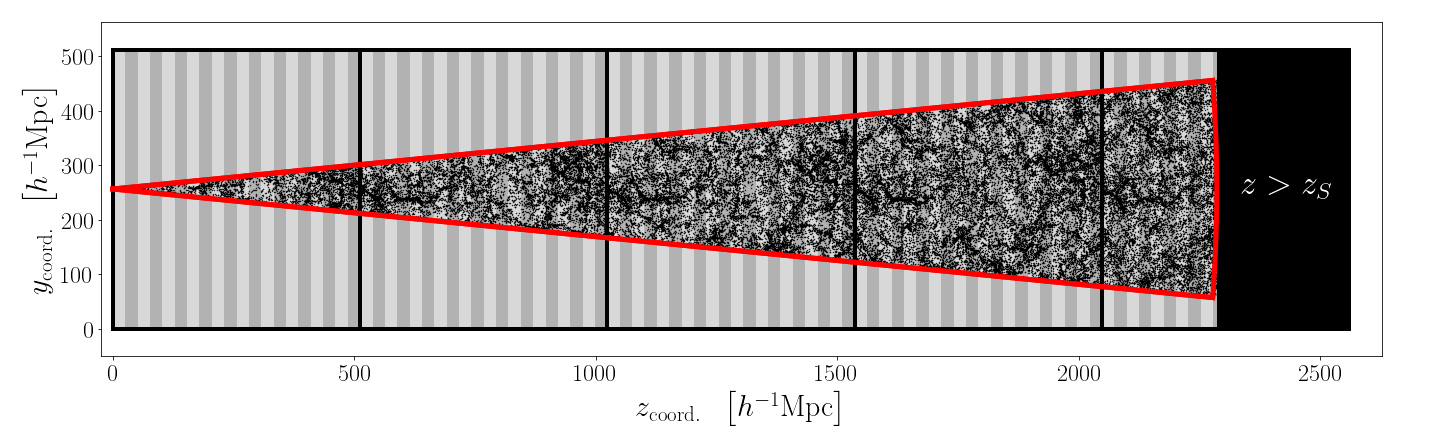}}
\caption{Sketch of the lightcone setup. The black solid lines represent the five $512 {\rm Mpc}/h$ simulation boxes that are tiled to encompass the $10\times10 {\rm deg}^2$ lightcone that extends out to $\zS = 1$. The lightcone geometry is depicted by the red solid line. The grey shaded slices mark the comoving distance (in the $(0,0,1)$ ($z_{\rm coord.}$) direction) covered by each of the 90 simulation snapshots. The comoving volume of the right-most simulation box that corresponds to $z>\zS$ (black) is never used in the calculations. In this work, all five simulation boxes correspond to the same $N$-body simulation, and hence, there is repetition of structure along the line-of-sight (this is however not important for our code comparison results).}
\label{fig:tiling}
\end{figure*}

We base our lensing simulation code comparison on a common $N$-body simulation of cosmic structure formation in a spatially flat standard cold-dark-matter cosmology with a cosmological constant. The assumed cosmological parameters are:
a matter density parameter $\OmegaMatter = 0.32$, a baryon density parameter $\OmegaBaryon = 0.049$, an amplitude parameter $\sigma_8 (z = 0) = 0.83$, spectral index $n_{\text{s}} = 0.96$ for the initial density power spectrum, and a Hubble constant $H_0 = h 100\,\kmsMpc$ with $h = 0.67$. In this work, we ignore the effects of the energy density of radiation, as well as massive neutrinos.

The simulation is carried out with the adaptive-mesh-refinement (AMR) code \ramses{} in a cubic box of side length $L = 512h^{-1}\,\Mpc$ with $1024^3$ matter tracer particles from $z = 99$ to redshift $z = 0$. The interpolation between the particle positions and the positions of the AMR mesh cells (needed to construct the density on the AMR grid and also evaluate the forces at the particle positions) is done with a cloud-in-cell (CIC) scheme. The cell refinement criterion adopted is $8$, i.e., an AMR cell is split into eight child-cells if the particle number in the cell exceeds 8. The initial conditions are generated with second-order Lagrangian perturbation theory using the routines implemented in the \pinocchio{} code \citep{2002MNRAS.331..587M,2017MNRAS.465.4658M}, which follow closely those of the initial conditions code \lptic{} \citep{1998MNRAS.299.1097S}. The initial conditions are generated using the linear power spectrum computed by the \camb{} code \citep{camb} at $z=0$, with its amplitude rescaled to $z=99$ using the linear growth factor of our fiducial cosmology. This ensures that the non-negligible contribution of radiation at $z=99$, that \camb{} captures but which we omit from the $N$-body simulation, does not cause discrepancies at the lower redshifts of interest.

In this work, we consider a single galaxy source redshift $\zS = 1$ and the particle information in the simulation is outputted at 90 redshift values between $z= \zS = 1$ and $z=0$. The output times are equally spaced in comoving distance $\chi(z)$ with intervals $\Delta\chi = 25.6h^{-1}\Mpc$. By tiling up these 90 simulation snapshots one can construct a realization of the time evolution of cosmic large-scale structure that encompasses a lightcone with $10\times10\ {\rm degree}^2$ out to $\chiS = \chi(\zS) \approx 2286h^{-1}\,\Mpc$. We place the observer at $(L/2, L/2, 0)$ in cartesian box coordinates, with the central line of sight of the observer's backward lightcone chosen along the $(0,0,1)$ direction. This lensing lightcone geometry is depicted in Fig.~\ref{fig:tiling}. 

All 90 snapshots are from the same $N$-body simulation, and as a result, the same large-scale structures appear at different epochs along the lightcone (most notably closer to the center of the field-of-view).
This issue can be avoided by running five different $N$-body simulations, each providing 20 snapshots along the lightcone (cf.~five black boxes drawn in Fig.~\ref{fig:tiling}).
When comparing results using just one $N$-body simulation to results using 5 simulations for tests, we do not detect any significant impact on our results, which focus on the differences between numerical methods for a fixed realization of the large-scale structure (with the exact degree of realism of such a realization being of secondary importance). We thus report only the results using a single $N$-body simulation for brevity.

We note also that our main goal is to assess the accuracy of the various lensing codes on their lensing predictions for small angular scales, for which our simulated field of view is sufficient. Our field of view is appreciably smaller than the expected area for Euclid ($100\ {\rm deg}^2$ vs. $15000\ {\rm deg}^2$), but our comparison analysis remains of interest for such wide-field imaging surveys since a significant portion of the constraining power comes from the smallest scales. Further, on larger angular scales, calculations based on linear theory are sufficient and the lensing codes tested here are not strictly needed to obtain theoretical predictions. 

\subsection{Lensing simulations}
\label{sec:simulations:lensing_simulations}

\begin{table*}
\centering
\begin{tabular}{@{}lccccccccccc}
\hline\hline
\rule{0pt}{1\normalbaselineskip}
Name &\ \ \hilbert{} & \ \ \fabbian{} & \ \ \mapsim{} & \ \ \mice{} & \ \ \rayramses{} 
\\
\hline
\rule{0pt}{1\normalbaselineskip}
Code paper & {\scriptsize \cite{2009AaA...499...31H}} & {\scriptsize \cite{fabbian2018}} & {\scriptsize \cite{2015MNRAS.452.2757G}} & {\scriptsize \cite{fosalba08}} & {\scriptsize \cite{2016JCAP...05..001B}}
\\
\\
\,\,Code type &  Post-process & Post-process  & Post-process  & Post-process & On the fly
\\
\,\, &  (multiple plane) & (multiple sphere)  & (multiple plane) & (multiple sphere) & 
\\
\\
\,\,LOS projection & $\parallel$ to central LOS & Radial & Radial & Radial & Radial
\\
\\
\,\,LOS resolution & Particle outputs & Particle outputs & Particle outputs & Particle outputs & {\sc RAMSES} time steps
\\
\\
\,\,Ray grid scheme & Regular grid & \healpix{}\footnotemark & Regular grid & \healpix{} & Regular grid 
\\
\\
\,\,Full-sky maps & w/ development & $\checkmark$  & w/ development & $\checkmark$ &  w/ development
\\
\\
\,\,Beyond-Born & $\checkmark$ & $\checkmark$ & w/ development & w/ development & w/ development 
\\
\\
\hline
\hline
\end{tabular}
\caption{Summary of the key features of the lensing simulation codes compared in this paper. The entries "w/ development" indicate that the code versions used do not immediately admit the corresponding feature, but that there is no impediment for it to be implemented with further development.}
\label{table:codes}
\end{table*}

We refer to the lensing simulation codes that participate in this comparison project as the \hilbert{} \citep{2007MNRAS.382..121H, 2009AaA...499...31H}, \fabbian{} \citep{fabbian2013, calabrese2015, fabbian2018}, \mapsim{} \citep{2015MNRAS.452.2757G}, \mice{} \citep{fosalba08,fosalba15} and \rayramses{}  \citep{2016JCAP...05..001B} codes. The main data produced by these codes for this analysis are lensing convergence maps with $2048^2$ pixels on a regular mesh that covers a $10\times10\,\degt^2$ field of view. This default pixel resolution corresponds to an angular resolution of $\approx 18\,\arcsect$ (due to the details of their operation, the \fabbian{} and \mice{} codes will work at slightly lower resolution).

The main goal of this paper is to evaluate the level of agreement between these maps for a number of different summary statistics. In the remainder of this section, we describe, in alphabetical order, the main aspects of the operation of these lensing simulation codes. We shall be succinct in the description and we refer the interested reader to the relevant cited literature for more details on their operation. Table \ref{table:codes} summarizes the main key features of the lensing simulation codes.

\subsubsection{\hilbert}
\label{sec:simulations:lensing_simulations:hilbert}

The \hilbert{} lensing simulation code is described in \citet[][]{2007MNRAS.382..121H,2009AaA...499...31H}. The code implements a multiple-lens-plane algorithm in the flat-sky approximation, and it is capable of computing convergence and shear fields both in full ray-tracing mode with multiple light deflections and lens-lens coupling, and within the Born approximation with unperturbed ray trajectories and without lens-lens coupling. 

In the first step of the code operation, the backward lightcone up to the source redshift $\zS = 1$ is divided into 90 redshift slices of $25.6h^{-1}\,\Mpc$ comoving thickness. Each slice is filled with the matter distribution of a snapshot of the $N$-body simulation. The matter of each slice is projected  parallel to the $(0,0,1)$ direction onto a lens plane located at the center of the slice. From the matter on the lens planes, the two-dimensional lensing potential and its first and second derivatives are computed using a particle-mesh particle-mesh (PMPM) method. On each lens plane, a coarse mesh with $16384^2$ mesh points and side length $512h^{-1}\,\Mpc$ spans the whole cross-section of the box. One or more finer meshes with $8192^2$ mesh points and $5h^{-1}\,\kpc$ mesh spacing are used to cover the intersection of the lightcone with each lens plane. The simulation particles are assigned to the two-dimensional mesh points using CIC assignment. The matter distribution on the meshes is smoothed further with a Gaussian kernel with a constant comoving width $\sigma_{\text{G}} = 10h^{-1}\,\kpc$ per dimension. Thus the smoothing is well captured by a kernel $\hat{W}$ in comoving harmonic space that is a product of the (circularly averaged) CIC mass assignment kernel and a Gaussian smoothing kernel:
\be\label{eq:hilbert_smoothing_kernel}
\hat{W}(k) =
\sinc\left(\frac{d_{\text{fine}} k}{2}\right)^4
\exp\left(-\frac{\sigma_{\text{G}}^2 k^2}{2}\right)
,
\ee
where $k$ denotes the comoving wave number, and $d_{\text{fine}} = 5h^{-1}\,\kpc$ denotes the fine mesh spacing.

The coarse meshes are used to compute a long-range low-pass filtered version of the lensing potential from the projected density using Fast Fourier Transforms (FFT). From that long-range potential, the first and second derivatives on the mesh points are computed using finite differencing. Similarly, the fine meshes are used to compute the complementary high-pass filtered short-ranged part of the lensing potential and its derivatives. 
 
Light rays are then traced back from the observer through the series of lens planes to the source plane. In full ray-tracing mode, deflection angles at each lens plane are computed by bilinear interpolation of the first derivatives of the lensing potential at the mesh points onto the ray position. The lensing Jacobians of the rays are updated using the second derivatives of the lensing potential. In Born mode, light rays are not deflected and the lensing Jacobians of the rays are computed by sums of the second derivatives with appropriate lensing efficiency weights.

\subsubsection{\fabbian}
\label{sec:simulations:lensing_simulations:fabbian}
\footnotetext{Despite adopting a \healpix{} grid for this work, \fabbian{} supports ray-tracing on arbitrary isolatitudinal grids on the sphere that are symmetric with respect to the equator. See \cite{fabbian2013} and references therein for more details.}
\fabbian{} implements a multiple lens ray-tracing algorithm in spherical coordinates on the full sky using the approaches of \citet{fosalba08} and \citet{dasbode2008}, and was originally developed to perform high-resolution CMB lensing simulations \citep{fabbian2013, fabbian2018}. The current version of the code reconstructs the full-sky backward lightcone around the observer using the particle snapshots produced by an $N$-body simulation out to the comoving distance of the highest redshift available from the snapshots following \cite{calabrese2015}. Because of the finite size of an $N$-body simulation box, the code replicates the box volume the necessary number of times in space to fill the entire observable volume between the observer and the source plane. In order to minimize systematics arising from the box replica (and thus by the repetition of the same structures along the line of sight) the code can randomize the particle positions as described in \cite{carbone2008, carbone2009}. However, this randomization is not performed here to ensure the \fabbian{} code \lq{}sees\rq{} the same large-scale structures as the other codes. Furthermore, the results reported here focus only of the $10\times10\ {\rm deg}^2$ portion of the sky that is common to all codes (cf.~Fig.~\ref{fig:tiling}). 

The backward lightcone is sliced into 90 full-sky spherical shells such that the median comoving distance spanned by each shell coincides with the comoving distance at the redshift of each $N$-body snapshot. The particles inside each of these shells are projected onto spheres. The surface mass density $\Sigma$ on each sphere is defined on a two-dimensional grid. For each pixel of the $i$-th sphere one has
\begin{equation}
\Sigma^{(i)}(\vtheta) = \dfrac{n m}{A_{\rm pix}}\,,
\label{eq:surfmass}
\end{equation}
where $n$ is the number of particles in the pixel, $A_{\rm pix}$ is the pixel area in steradians and $m$ is the particle mass of the $N$-body simulation in each pixel. 

For this work, the \fabbian{} code produces a full-sky convergence map on a \healpix{}\footnote{\url{http://healpix.sourceforge.net}} grid \citep{HEALPix} with $N_\text{side} = 8192$, which corresponds to a pixel resolution of $26\,\arcsect$. 
The lensing convergence of a source plane at redshift $\zS$ in the Born approximation is computed as the lensing-efficiency-weighted sum of the surface mass density:
\be\label{eq:kappa_fabbian}
\begin{split}
 \kappa (\vtheta, \chiS) &= 
\frac{4 \pi \GNewton}{\clight^2}\frac{1}{\fS}\!
\sum_{i} 
(1 + \zD^{(i)}) \frac{\fDS^{(i)}}{\fD^{(i)}}
\left[\Sigma^{(i)}(\vtheta)\!-\!\bar{\Sigma}^{(i)}\right]
,
\end{split}
\ee
where $\Sigma^{(i)}$ denotes the angular surface mass density, $\bar{\Sigma}^{(i)}$ is the mean angular surface mass density of the $i$-th shell, and $\fDS^{(i)}$ and $\fD^{(i)}$ are the corresponding distances at the redshift of the $i$-th shell. The angular position of the center of each \healpix{} pixel coincides with the direction of propagation of the rays in the Born approximation.

The common sky patch for the code comparison is extracted from the \healpix{} map with a Lambert azimuthal equal area projection using $18\,\arcsect$ pixels\footnote{We use the \texttt{azeqview} routine of the \healpix{} python package \healpy{}.}. 
We correct for the effect of projecting the \healpix{} map to the higher-resolution flat-sky map in the convergence power spectrum estimation by multiplying $\harmPS{}{\kappa}{\ell}$ by a pixel window function $w_{\ell}^2$ estimated as follows. We first synthesize 100 Gaussian realizations of a convergence field on a $\nside=8192$ \healpix{} grid from a theoretical power spectrum $\harmPS{\text{MC}}{\kappa}{\ell}$ at $\zS=1$ using the \healpix{} \texttt{synfast} routine. These realizations are then similarly projected to a flat-sky and their power spectrum measured. The pixel window function $w_{\ell}$ is then defined as 
\begin{equation}
w_{\ell}^{-2} = \frac{\langle \harmPS{\text{MC}, {\rm proj.}}{\kappa}{\ell} \rangle}{\harmPS{\text{MC}}{\kappa}{\ell}},
\end{equation}
where the angular brackets denote the average over the power spectrum measured from each projected Gaussian field $\harmPS{\text{MC}, {\rm proj.}}{\kappa}{\ell}$.

\fabbian{} can also be used to propagate the lensing Jacobian beyond the Born-approximation \citep{fabbian2018}. In this case the two-dimensional lensing potential and its derivatives required by the multiple-lens algorithm are computed in the harmonic domain by solving the Poisson equation, and later resampled on a higher-resolution ECP pixelization \citep{ecp} that can reach the $\arcsect$ resolution. The perturbed ray trajectories are then computed using a nearest grid point interpolation scheme.

\subsubsection{\mapsim}
\label{sec:simulations:lensing_simulations:mapsim}

The \mapsim{} code has been developed by \cite{2015MNRAS.452.2757G} and it works in two main steps termed i-\mapsim{} and ray-\mapsim{}.  In the first step  i-\mapsim{}, the particle positions in the simulation snapshots that lie within the desired field-of-view are projected onto different lens planes located along the line-of-sight. Each particle is placed in the nearest lens plane maintaining angular positions. The mass density is then interpolated from the projected particle positions to a two-dimensional grid using a triangular shaped cloud (TSC) scheme. The grid pixels are chosen to have the same angular size on all lens planes and no particle randomization is performed to ensure \mapsim{} calculates the lensing signal using the same large-scale structure as the other codes. The angular surface mass density $\Sigma^{(i)}$ on the $i$-th plane is computed as in Eq.~\eqref{eq:surfmass}.

In the second step \citep[as done by][]{petri16,petri17,giocoli17,giocoli18a,  castro17}\footnote{
The mass maps produced in this first step i-\mapsim{} can also be used as input to the \glamer{} lensing code \citep{2014MNRAS.445.1942M, 2014MNRAS.445.1954P} to perform ray-tracing calculations (including beyond the Born approximation).
}, ray-\mapsim{} constructs the lensing convergence map in the Born approximation by simply summing up the surface mass density from each plane along the line-of-sight, weighted appropriately by the lensing efficiency kernels as in Eq.~\eqref{eq:kappa_fabbian}, except now $i$ labels planes perpendicular to the $(0,0,1)$ direction instead of concentric spheres.
Since the grid pixels have the same angular resolution, they are also the direction of propagation of the light rays in the Born approximation. There is thus no need to interpolate the projected density defined on the grid to the exact light ray position.

\subsubsection{\mice}
\label{sec:simulations:lensing_simulations:fosalba}

The methodology of the \mice{} code presented in \cite{fosalba08, fosalba15} guided the development of the \fabbian{}, and hence, the two codes work very similarly. The $N$-body data is sliced and projected onto spherical shells \citep[called the "onion universe" in][]{fosalba08}. Each such shell is used to define surface mass density fields on \healpix{} grids. Finally, the surface mass is summed along the line-of-sight, weighted by the appropriate weak lensing efficiency factors. 

The \mice{} lensing maps produced for this comparison project take the \healpix{} data from \fabbian{} as input. The lensing convergence is then calculated \emph{independently of} \fabbian{} employing the Born approximation. The common sky patch is extracted from the \healpix{} map with a Lambert azimuthal equal area projection using $18\,\arcsect$ pixels, and the power spectra measured from the patch are corrected for the projection. 

\subsubsection{\rayramses}
\label{sec:simulations:lensing_simulations:rayramses}

The \rayramses{} code is described in detail in \citet{2016JCAP...05..001B}. The computation of the lensing quantities is based on the original ideas of \cite{2000ApJ...537....1W, 2011MNRAS.415..881L} and it is done on-the-fly during the \ramses{} $N$-body simulation that produced the 90 snapshots that serve as input to the other codes. This code therefore does not rely on any discretization of the density field along the line-of-sight, or more precisely, it retains the full line-of-sight (or time) resolution attained by the $N$-body simulation itself. Likewise, \rayramses{} also bypasses the need to choose a density assignment scheme to construct two-dimensional projected density planes from the three-dimensional density distribution, i.e., the transverse spatial resolution is directly specified by the AMR grid structure of the \ramses{} simulation. 

In \rayramses{}, light rays are initialized at the curved surface of constant $\zS=1$ and then subsequently followed in unperturbed trajectories until they reach the observer at $z=0$. In each simulation particle time step, the rays are moved by the distance light would travel during that time interval. The size of the time steps depends on the refinement of the grid in \ramses{}: rays located in refined regions are integrated more often than rays in unrefined regions. The lensing integral accumulated during a time step is the sum of the lensing integral associated with each crossed cell:\footnote{This is integration \lq{}method B\rq{} in \citet{2016JCAP...05..001B}.} 
\begin{equation}
 \kappa_{\rm ts} =
  \frac{1}{\clight^2}
 \sum_{\text{cells}} 
  \int_{\chiD^{\text{cell,end}}}^{\chiD^{\text{cell,start}}} \diff \chiD
  \frac{\fD\fDS}{\fS} \nabla^2_{\text{2D}} \Phi,
\label{eq:rayramses1}
\end{equation}
where $\chiD^{\text{cell,start}}$ and $\chiD^{\text{cell,end}}$ are the comoving distances between the observer and ray at the start and end of its trajectory inside a given mesh cell, and the sum runs over all of the cells crossed by the ray during the time step.  The total lensing convergence is the sum of the convergence accumulated during all of the simulation time steps $\kappa = \sum_{ts} \kappa_{\rm ts}$.

The quantity $\nabla^2_{\text{2D}} \Phi = \nabla_1\nabla^1\Phi + \nabla_2\nabla^2\Phi$ is the two-dimensional Laplacian of the gravitational potential ($\nabla_1$ and $\nabla_2$ represent the curved-sky generalizations of $\partial/\partial\theta_1$ and $\partial/\partial\theta_2$ in Sec.~\ref{sec:theory:wl}), which is related to the three-dimensional second-derivative $\nabla_i\nabla_j \Phi$ ($i,j = x,y,z$) via geometrical factors determined by the direction of motion of the ray. The values of $\nabla_i\nabla_j \Phi$ are evaluated by finite-differencing the potential in neighbouring cells, analogously to how standard \ramses{} computes the force. In the calculations presented in this paper, the values of $\nabla_i\nabla_j \Phi$ are treated as constant inside each cell.\footnote{\rayramses{} allows also to evaluate $\nabla_i\nabla_j \Phi$ inside each cell via trilinear interpolation using the quantity's value reconstructed at the cell vertices \citep[see][for the details]{2016JCAP...05..001B}.} \rayramses{} can evaluate also the shear components $\gamma_1$ and $\gamma_2$ by replacing $\nabla^2_{\text{2D}} \Phi$ in Eq.~(\ref{eq:rayramses1}) with $\nabla_1\nabla^1\Phi - \nabla_2\nabla^2\Phi$ and $2\nabla_1\nabla^2\Phi$, respectively. The calculation of these quantities for a given cell involves the information from a larger number of neighbouring cells compared to the calculation of $\nabla^2_{\text{2D}} \Phi$ for $\kappa$. This constitutes an effective form of smoothing that explains some differences between the convergence and shear power spectra of \rayramses{}. We will return to this point below in Sec.~\ref{sec:results:shear}.

As it runs on-the-fly with the simulation, the generation of lensing maps using \rayramses{} with certain specifications modified (e.g. increased number of pixels) requires rerunning the $N$-body simulation. All of the \rayramses{} results shown below correspond to the default $2048^2$ ($\approx 18\,\arcsect$) map resolution. 

\section{Results}
\label{sec:results} 

In this section, we compare the various lensing codes by analysing a number of statistics of the lensing maps: their probability distribution function (PDF), the convergence power spectrum and lensing peak counts. We also study the impact that a number of variations in code setups (e.g. Born vs. beyond-Born approximation, smoothing schemes, line-of-sight resolution, convergence vs. shear power spectra) can have on the results. 

\subsection{Convergence maps}
\label{sec:results:convergence_maps}

\begin{figure*}
\centerline{\includegraphics[width=\linewidth]{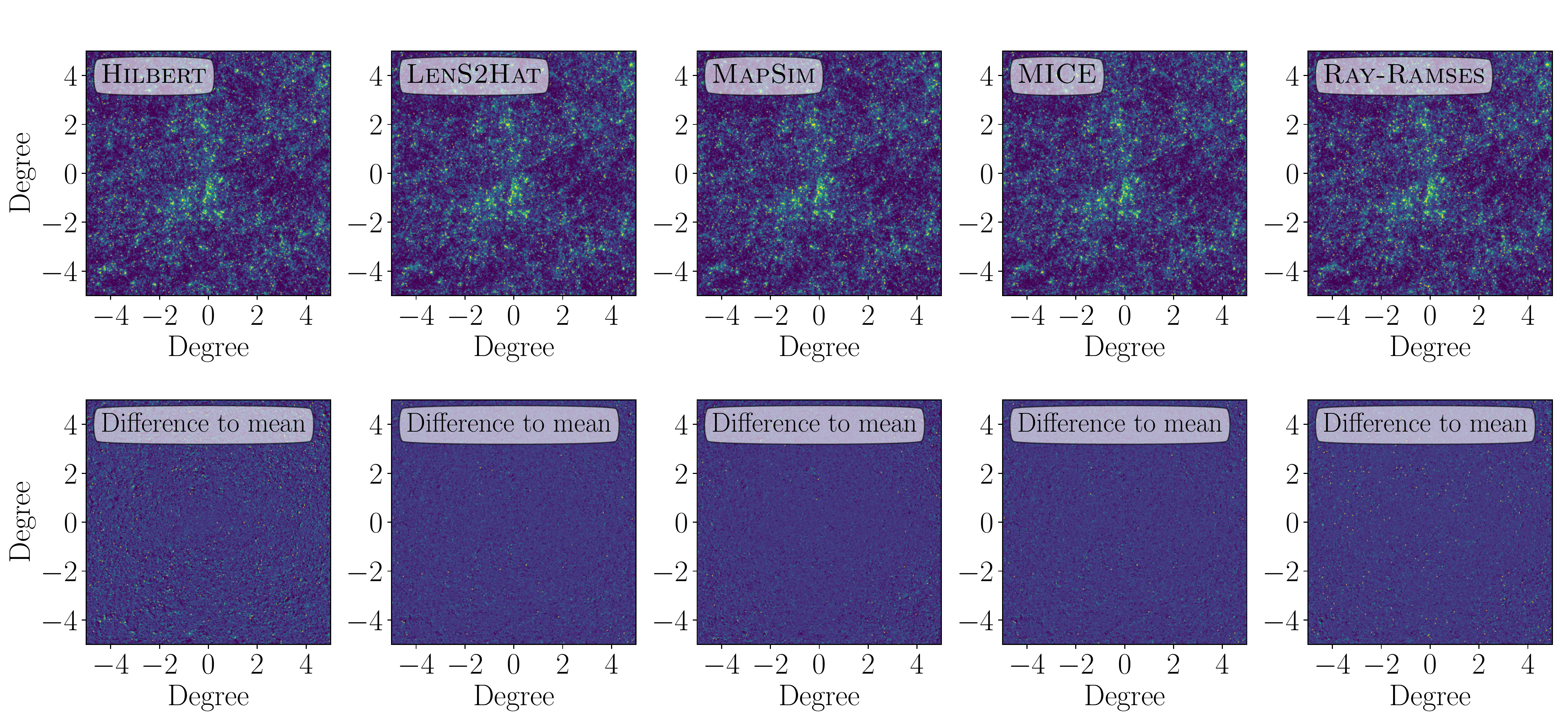}}
\caption{Lensing convergence maps produced by the lensing simulation codes, as labeled ($10\times10\ {\rm deg}^2$ with $2048\times2048$ resolution). The upper panels show the convergence maps as obtained by the codes. The lower panels show the corresponding difference to the mean of the codes. The color scale is the same in all panels, ranging from $\kappa = -0.02$ (dark blue) to $\kappa = 0.1$ (bright yellow).}
\label{fig:maps}
\end{figure*}

\begin{figure*}
\centerline{\includegraphics[width=\linewidth]{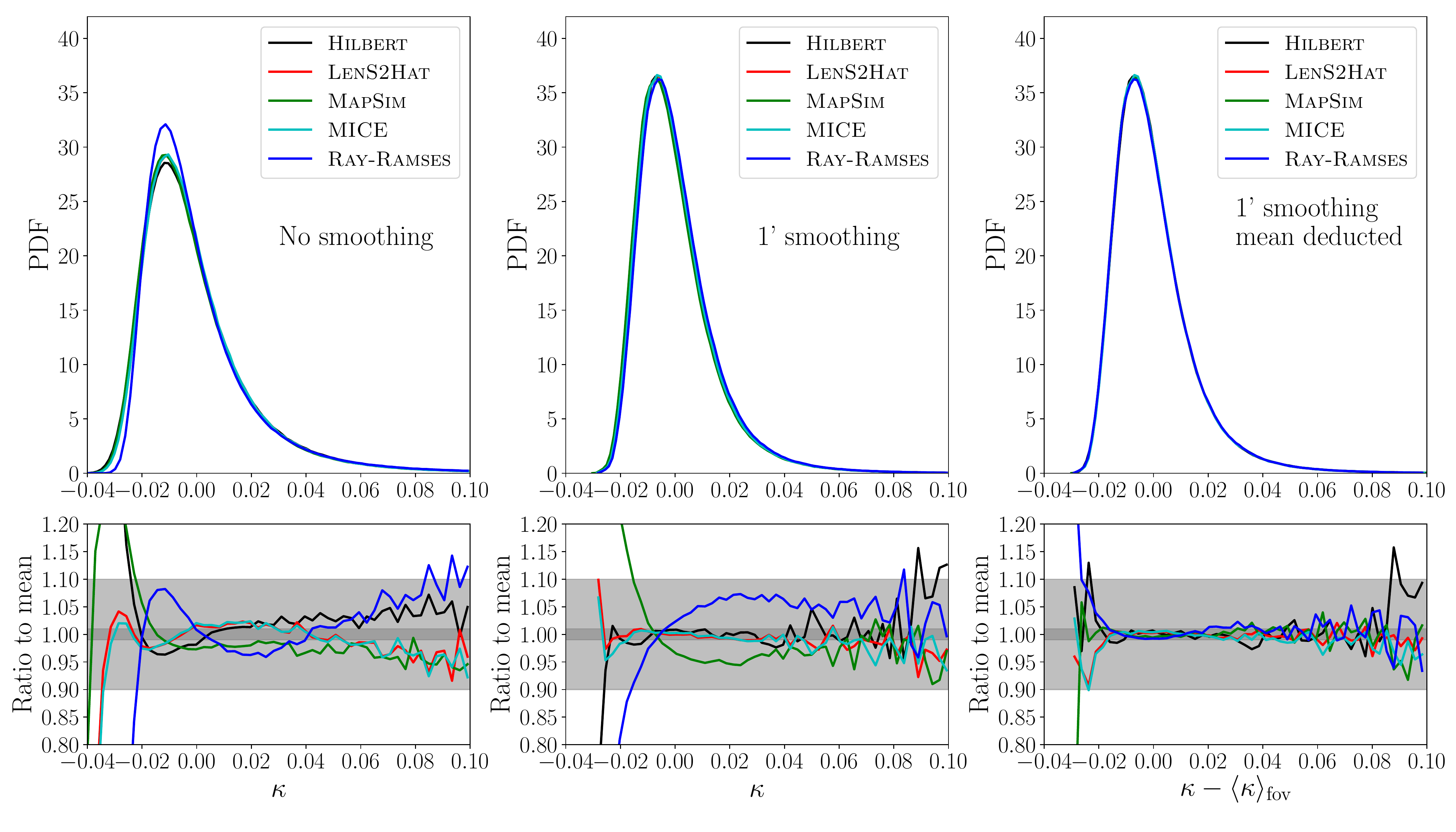}}
\caption{Probability density functions (PDF) of the convergence maps. The upper panels show the PDFs obtained by the lensing codes without smoothing (left), after smoothing using a Gaussian kernel with width $1\ \arcmint$ (center), and after smoothing and deducting the field mean value (right). The lower panels show the corresponding difference to the PDF given by the mean of the codes. The light and dark gray shaded areas indicate $10\%$ and $1\%$ fractional errors. In all of the upper panels, the read and cyan lines are practically indistinguishable; in the right upper panel, all curves are indistinguishable.}
\label{fig:pdfs1} 
\end{figure*}

The convergence maps produced by the different lensing simulation codes are compared in Fig.~\ref{fig:maps}. This comparison serves as a basic sanity check that the codes were successfully run on the same lightcone geometry and cosmic large-scale structure. This is confirmed by the good agreement between the position of high-$\kappa$ and low-$\kappa$ regions, as well as the small differences to the mean map of the codes.

The convergence one-point distributions of the maps are shown in Fig.~\ref{fig:pdfs1}. In the absence of any smoothing performed on the maps (left panels), the PDF of the convergence from \rayramses{} has a noticeably higher peak at $\kappa \approx -0.01$, and it decays more sharply towards more negative $\kappa$ values. The \rayramses{} PDF also has higher probability of large $\kappa > 0.06$ values than the other codes. This indicates a stronger smoothing of the matter by \rayramses{} in low-density regions and a higher resolution in high-density regions compared to the other codes. This is as one would expect from the underlying adaptive three-dimensional mesh employed in \rayramses{} compared to the non-adaptive two-dimensional grids of the other codes.

After smoothing the maps with a Gaussian kernel with size $1\,\arcmint$ (center panels), the shapes of the PDFs of all five codes are brought closer together, but slight horizontal shifts in the corresponding curves remain. This indicates that the codes produce slightly different mean values of the convergence across the field of view $\langle\kappa\rangle_{\rm fov}$. When the mean is taken out (right panels), the PDFs agree to $\approx 1\%$ in the range of $\kappa$ values where the PDFs are sizeable. We do not investigate further the origin of remaining differences on the tail of the distribution given the many different details in the operation of the codes. For example, the differences in the exact way the projection onto discrete planes/spheres is done in the codes (projection along line-of-sight vs. along $z_{\rm coord}$, different density assignment schemes, etc.) can cause small differences that would appear exacerbated in relative differences of a (small) PDF. We note also that the slight difference in the values of $\langle\kappa\rangle_{\rm fov}$ is not worrying as it has little impact on lensing observables. For example, in the remainder of this section, we compare statistics measured from maps without the mean subtracted and we will find very good agreement between the codes. Furthermore, analyses based on shear statistics are more closely related to the actually observed galaxy ellipticities and are not sensitive to shifts in the mean signal across the field of view.

\subsection{Convergence power spectra}
\label{sec:results:convergence_power_spectra}

\begin{figure}
\centerline{\includegraphics[width=\linewidth]{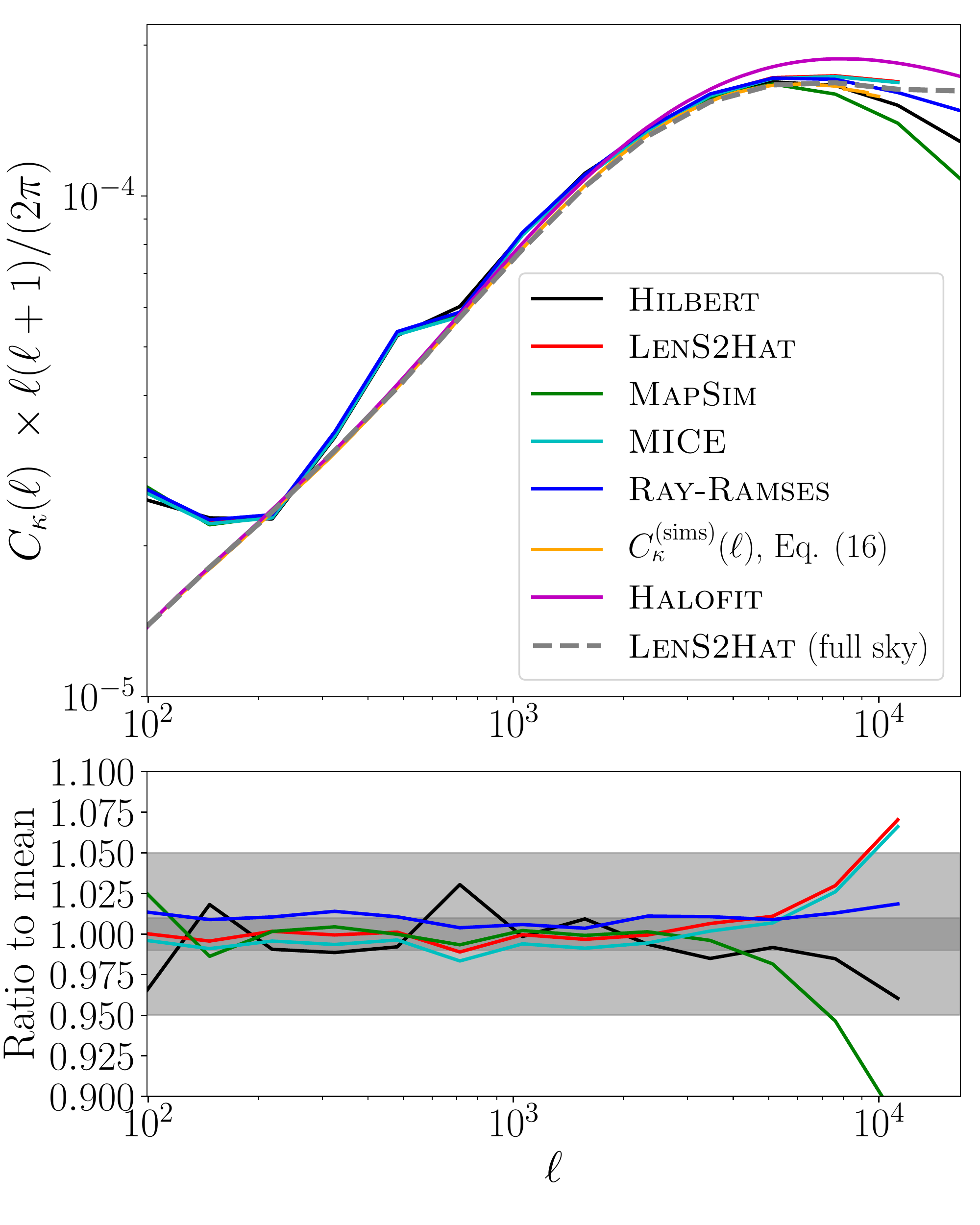}}
\caption{Power spectrum of the lensing convergence maps obtained by the lensing simulation codes, as labeled (upper panel). The result obtained by integrating the three-dimensional nonlinear matter power spectrum measured in the simulations according to Eq.~\eqref{eq:limberPk} is also shown for reference (orange line). The magenta curve shows the convergence power spectrum obtained by integrating the matter power spectrum given by the \halofit{} formula. In the upper panel, the red and cyan lines are nearly indistinguishable. The lower panel shows the ratio to the mean of the codes. The light and dark gray shaded areas indicate $5\%$ and $1\%$ fractional errors.}
\label{fig:spectra}
\end{figure}

The power spectra of the convergence maps of the \hilbert{}, \fabbian{}, \mapsim{}, \mice{} and \rayramses{} codes are shown in Fig.~\ref{fig:spectra}. The power spectra are calculated with Fourier transforms assuming the flat-sky approximation, which is valid for the small field-of-view our comparison is based on.\footnote{In cases where the curvature of the sky cannot be ignored, the spectra need to be computed using spherical harmonic decompositions instead.}
All spectra were evaluated using the routines in the publicly available \lenstools{} software \citep{2016A&C....17...73P}.\footnote{
We checked that our results are identical using an independent power spectrum calculation code.}
 The resulting spectra are then subsequently averaged in logarithmically spaced $\ell$-bins.
Figure~\ref{fig:spectra} also shows the result obtained by integrating the three-dimensional matter power spectrum measured directly from the simulation snapshots according to Eq.~\eqref{eq:limberPk}, as well as integrating the power spectrum given by the \halofit{} fitting formula \citep{2003MNRAS.341.1311S, 2012ApJ...761..152T}. 

The lower panel of Fig.~\ref{fig:spectra} shows the ratio of the individual code results to their mean. The \fabbian{}, \mapsim{}, \mice{}, and \rayramses{} codes agree with the mean of the codes to $\lesssim 2\%$ on scales $\ell \lesssim 4000$. The same holds for the \hilbert{} code, when ignoring the fluctuations of $\approx 3\%$ at $\ell \approx 100$ and $\ell \approx 700$. These larger differences are likely due to the parallel projection (i.e. along the $z_{\rm coord.}$) used by \hilbert{} in contrast to the radial projection employed by the other codes, which causes slight differences in what structures appear where in the field of view. Note also that on scales $\ell \gtrsim 4000$, where the code differences become larger ($10\%$ for $\ell = 10^4$ in some cases), uncertainties associated with the modeling of baryonic processes (most notably stellar and AGN feedback) are expected to be sources of larger systematic errors \citep{2011MNRAS.417.2020S, 2019arXiv190402070B, 2019MNRAS.tmp.1672H, 2019arXiv190407905G}. Overall, we therefore conclude that the lensing codes tested display a satisfactory level of agreement, which is a valuable cross-check in preparation for the analysis of future surveys. 

On scales $\ell \gtrsim 3000$, the codes systematically underpredict the amplitude of the power spectrum predicted by the \halofit{} result. This reflects the lower resolution of the $N$-body simulation used to construct the convergence maps of the tested codes. This is confirmed by the fact that the result of Eq.~\eqref{eq:limberPk}, which uses the power spectrum directly measured from the snapshots, displays the same level of suppression on small scales relative to the \halofit{} curve. 

The code results also differ from Eq.~\eqref{eq:limberPk} on scales $\ell \lesssim 1000$. On these angular scales, the results are severely affected by sample variance as they correspond to a single realization of a rather small field of view; it is in fact a reassuring cross-check that the codes all agree on this peculiar shape of the spectra. Increasing the size of the field of view and/or the number of realizations of our $10\times10\ {\rm deg}^2$ field of view (and then taking the mean) would reduce the impact of sample variance. As a sanity check, the gray dashed line shows the spectra measured from the full-sky map produced by the \fabbian{} code, i.e., without specifying to the common $10\times10\ {\rm deg}^2$ field of view. This effectively increases the number of modes sampled (though not all independent because of box replication), which brings the result closer to the large-scale prediction of Eq.~\eqref{eq:limberPk}.

As an additional test, we also investigate the power spectrum of the log-transformed convergence \citep{2009ApJ...698L..90N, 2011ApJ...729L..11S, 2012ApJ...748...57S, 2013ApJ...763L..14M, 2017MNRAS.472L..80L}, which is sensitive to higher-order statistics of the convergence field itself. We find that the power spectra of ${\rm log}_{\kappa}\left(0.1 + \kappa\right)$  (the value of $0.1$ ensures the argument is always positive for our maps) measured from the maps of the codes agree with their mean to better than $\approx 2\%$ on scales $\ell \lesssim 3000$. 

\subsection{Peak counts}
\label{sec:results:peak_counts}

\begin{figure}
\centerline{\includegraphics[width=\linewidth]{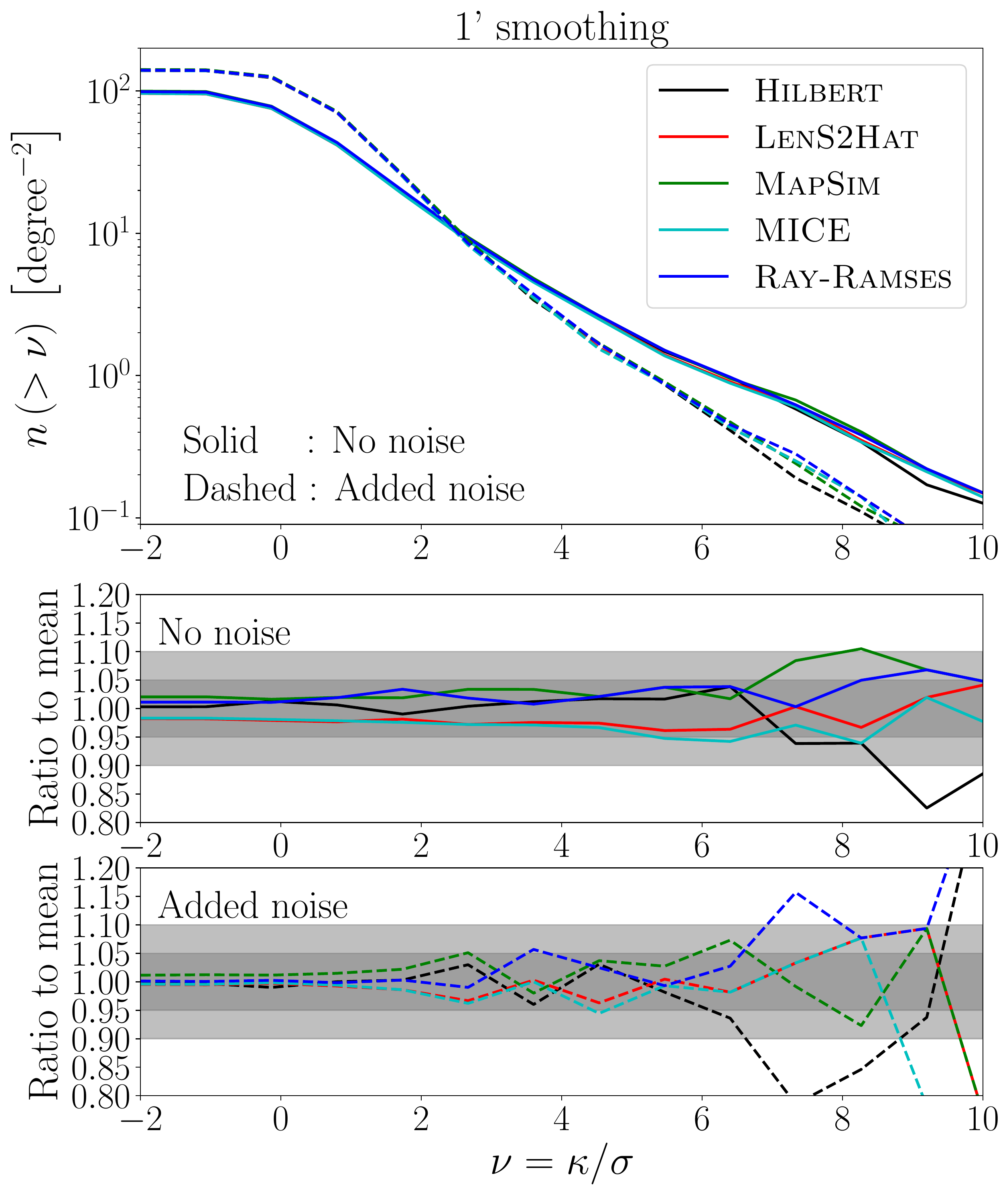}}
\caption{Cumulative count of lensing convergence peaks. The result is shown for the maps obtained by the various codes, with (dashed) and without (solid) shape noise added to the maps, as labeled. The result shown is for maps smoothed with a Gaussian kernel with the size of $1$ arcmin. The lower panels show the ratio to the mean result of the codes. The light and dark gray shaded areas indicate $10\%$ and $5\%$ fractional errors.}
\label{fig:peaks}
\end{figure}

Lensing peak counts contain information beyond the power spectrum, and their inclusion in data analyses has been advocated to be able to yield significantly improved cosmological parameter constraints \citep{2010MNRAS.402.1049D,2012MNRAS.423.1711M,2012MNRAS.426.2870H,2015PhRvD..91f3507L,2016PhRvL.117e1101L,2018MNRAS.474.1116S, 2019arXiv190501710D}. Figure \ref{fig:peaks} shows the cumulative number density of lensing convergence peaks found in the maps of the different codes. Peaks are identified as pixels whose amplitude is higher than that of all its 8 neighbours and the result is plotted in terms of the peak height (or signal-to-noise) $\nu = \kappa/\sigma$, where $\sigma$ is the standard deviation of all pixels. We use the $\kappa$ and $\sigma$ values of each code to define their peak height. We also count peaks on maps with their mean convergence subtracted (cf.~Sec.~\ref{sec:results:convergence_maps}). The result shown is for peaks counted on maps smoothed with a Gaussian filter of size $1\ {\rm arcmin}$, and with and without Gaussian shape noise added (assuming a variance $\sigma_e^2 = 0.31^2$ for the shape noise in the shear estimate per galaxy and an effective number density of source galaxies $\bar{n}_{\rm eff} = 30\,\arcmint^{-2}$; the values of both $\kappa$ and $\sigma$ in $\nu = \kappa/\sigma$ are computed for maps with and without added noise). The assumption of Gaussian distributed shape noise is not critical to the code comparison.

The codes all display very similar lensing peak count predictions. In particular, for both the cases with and without shape noise, the codes agree with the mean prediction to better than $5\%$ for peaks with $\nu \lesssim 6$. This level of agreement is naturally related to that observed in Sec.~\ref{sec:results:convergence_maps} for the convergence maps PDFs. For higher $\nu$, the relative differences increase as the cumulative peak count becomes more sensitive to small changes. Should current or future real data analyses require better than $\sim 20\%$ precision for $\nu \sim 10$, then these code differences should be understood better. 

For completeness, we have also compared the codes for two additional cases (not shown): (i) peak counts on maps without the mean convergence deducted; (ii) using the values of $\sigma$ found in the map of one of the codes to define $\nu$ for all codes. We found that using a common $\sigma$ value yields effectively the same level of agreement and skipping subtracting the mean convergence yields an agreement of $10\%$ for $\nu \lesssim 6$.

\subsection{Systematic errors}
\label{sec:results:systematic_errors}

In this section, we quantify various possible sources of systematic errors in lensing simulations. 
Unless specified otherwise, the results described here are produced by the \hilbert{} code, and correspond to the average over 16 realizations of the $10\times10\,\degt^2$ field-of-view, obtained by picking 16 different observer orientations in the simulation box. This reduces the statistical noise due to the finiteness of the field of view (by up to a factor of four, assuming the realizations are statistically independent). 

\subsubsection{Born approximation}
\label{sec:results:systematic_errors:power_spectra:born_approximation}

\begin{figure}
\centerline{\includegraphics[width=\linewidth]{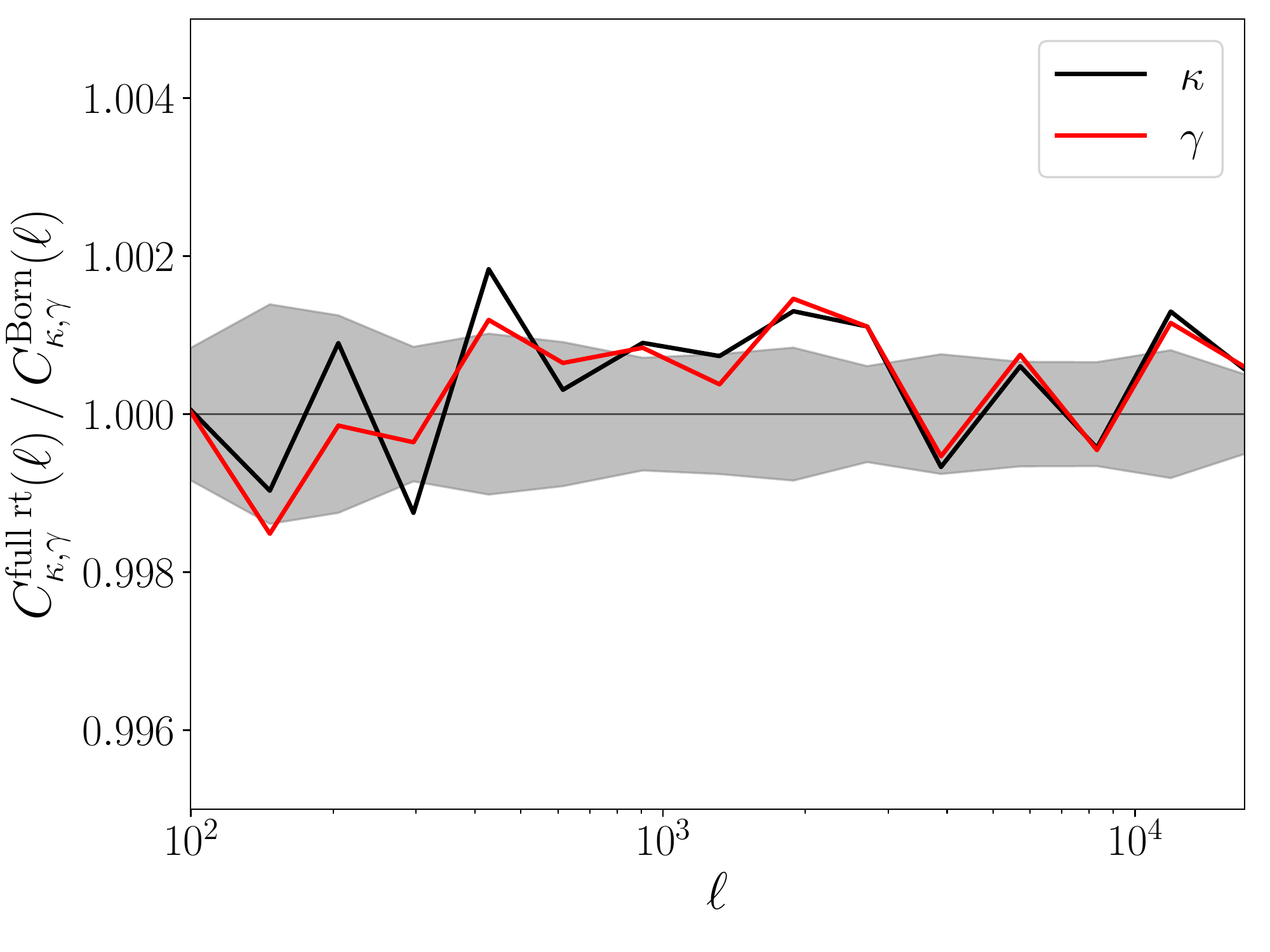}}
\caption{Ratio of the convergence and shear power spectra from the \hilbert{} code run with full ray-tracing to that assuming the Born-approximation. The lines show ratios of the mean over 16 $10\times10\,\degt^2$ fields. The grey region indicates the statistical error on these ratios (in terms of standard deviation) estimated from the field-to-field variation.}
\label{fig:born_vs_full_ray_tracing}
\end{figure}

As mentioned in Sec.~\ref{sec:theory:wl}, the impact of adopting the Born approximation (i.e. computing the lensing signal along unperturbed ray trajectories) on the convergence or shear power spectrum is expected to lie comfortably below the $1\%$ level for angular scales relevant for current and future surveys. This can be explicitly checked by comparing full beyond-Born ray-tracing with Born-approximation results. Conversely, assuming that the Born approximation is valid to better than $1\%$ on a given range of angular scales, such a comparison can also be used as a self-consistency check of a lensing simulation code that can perform both types of calculations. 

Figure \ref{fig:born_vs_full_ray_tracing} shows the ratio of the convergence and shear power spectra obtained without adopting the Born approximation to those obtained adopting it. The relative differences are $\sim 0.1\,\%$ out to $\ell = 10^4$ and are likely due to numerical noise\footnote{The main source of noise is that, for our finite fields of view, the lightcone in full-raytracing contains slightly different matter structures than the one in Born mode due to the presence/absense of light deflections.} rather than a direct consequence of the Born approximation. This corroborates past similar conclusions in the literature \citep[e.g.][]{2000ApJ...530..547J,2009AaA...499...31H,giocoli16a, fabbian2019}, but recall, as noted in Sec.~\ref{sec:theory:wl}, the degree of validity of the Born-approximation can be different for other lensing statistics.

\subsubsection{Line-of-sight discretization}
\label{sec:results:systematic_errors:line_of_sight_discretization}

\begin{figure}
\centerline{\includegraphics[width=\linewidth]{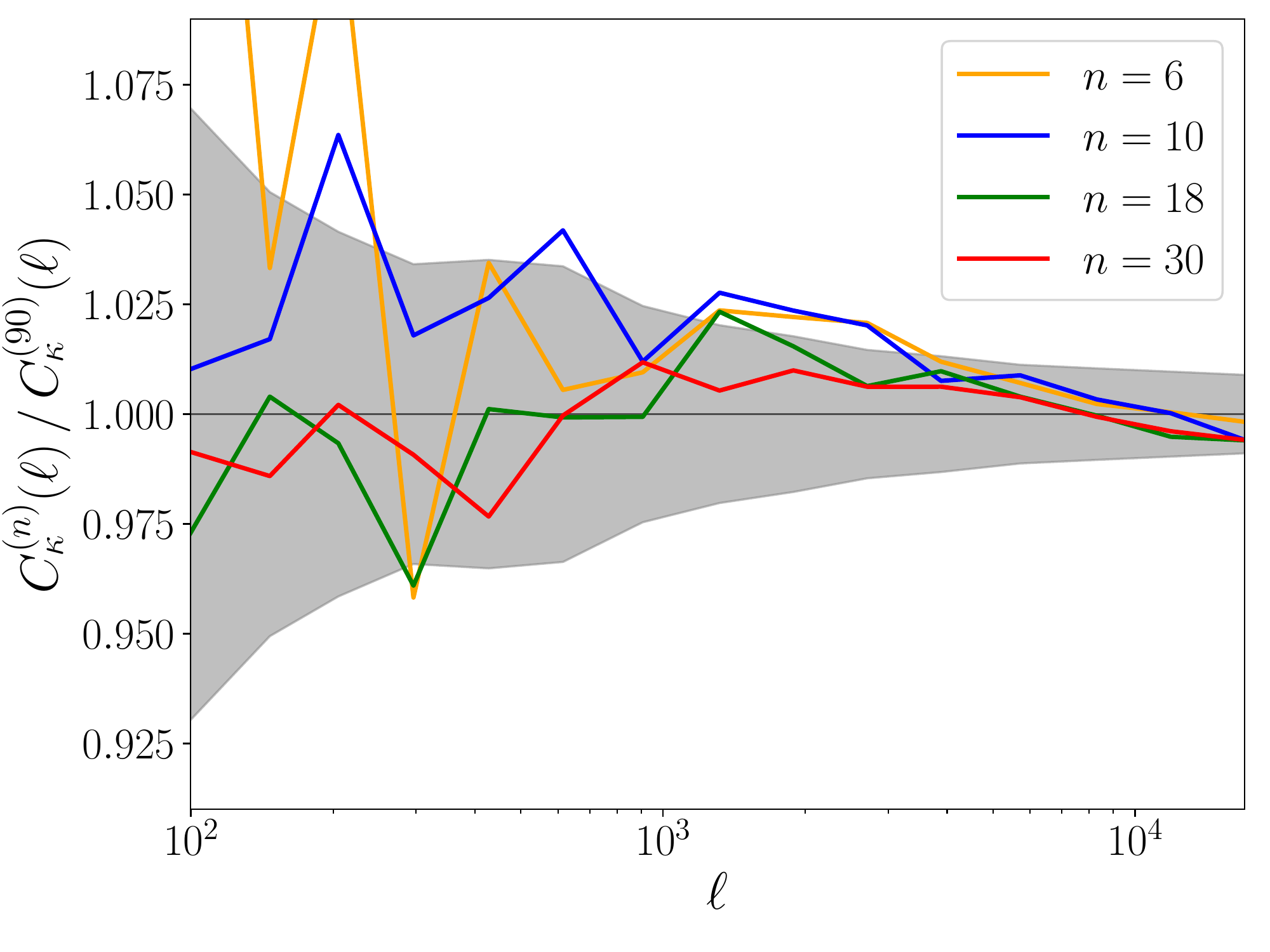}}
\caption{Ratio of the convergence power spectra $\harmPS{(n)}{\kappa}{\ell}$ for different number $n$ of planes to that using $n =  90$ planes for the \hilbert{} code. The results shown correspond to the mean of the 16 $10\times10\,\deg^2$ fields. The grey region indicates the relative statistical error (in terms of standard deviation) on the mean of the convergence power spectrum measured from these fields for the $n=90$ case.}
\label{fig:number_of_planes}
\end{figure}

\begin{figure}
\centerline{\includegraphics[width=\linewidth]{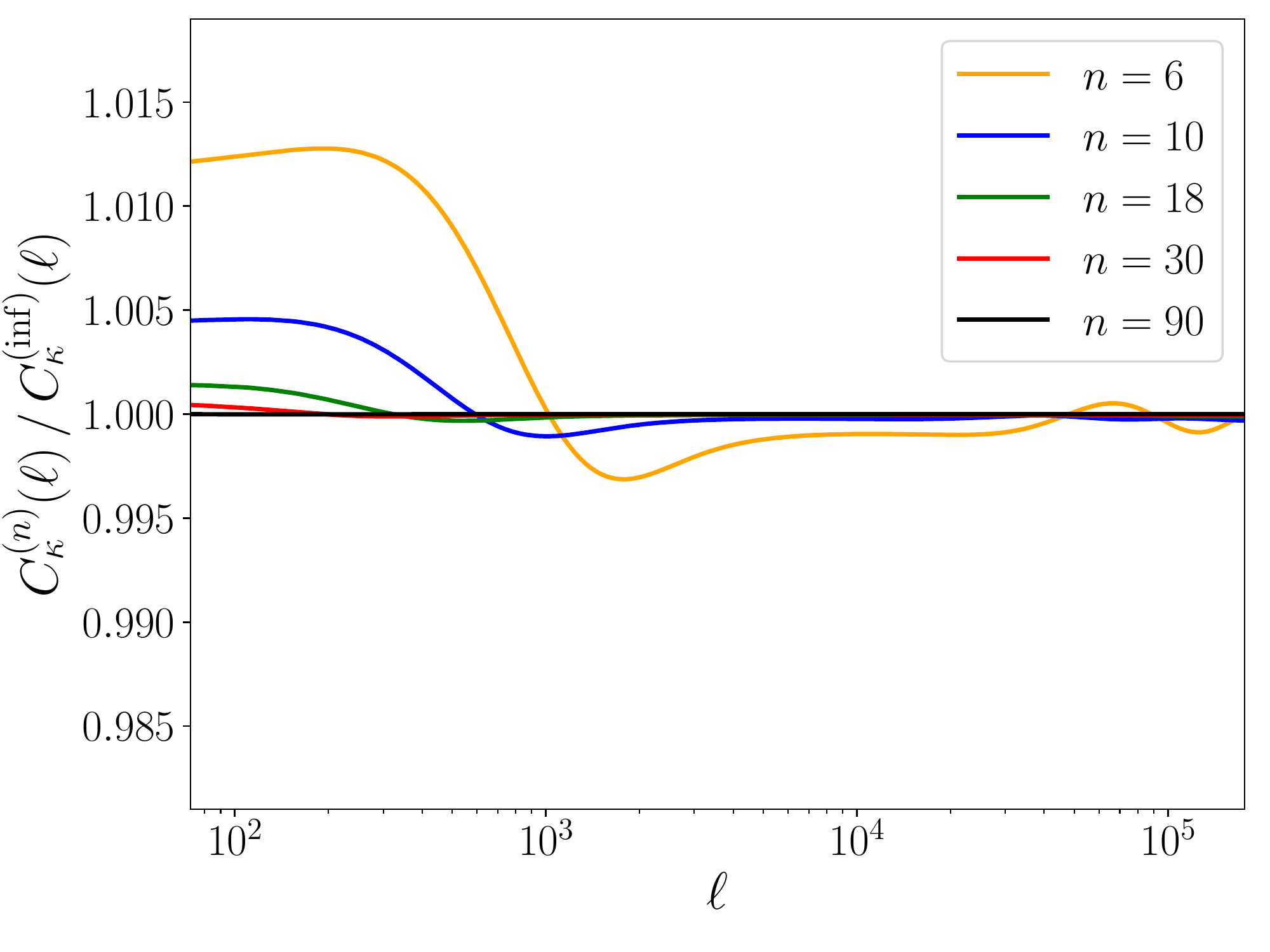}}
\caption{Ratio of the convergence power spectra $\harmPS{(n)}{\kappa}{\ell}$ for a finite number $n$ of planes to that in the limit $n\to \infty$, according to Eq.~\eqref{eq:limberPk_planes_and_smoothing}.}
\label{fig:number_of_planes_th}
\end{figure}

The users of multiple-plane and multiple-sphere algorithms must decide how many planes or spheres the algorithm employs to represent the matter distribution along the line-of-sight. Employing more planes/spheres naturally incurs on higher computational, data storage, and data transfer costs. Fewer planes/spheres may lead to larger line-of-sight discretization errors.

Figure \ref{fig:number_of_planes} illustrates for the \hilbert{} code, how the number of planes affects the resulting convergence power spectra. Reducing the number of planes from 90 to 30 or even to 10 may cause deviations in the measured power by up to 5\%.
This is in rough accordance with earlier results by \citet{2003ApJ...592..699V}. However, the differences we find are within the expected statistical error for such a finite field, and bear no strong indication for a systematic difference. Further, recall that the \hilbert{} code projects the matter onto planes along the $(0,0,1)$ direction, which exacerbates the impact of reducing the number of planes, compared to the other codes which do radial projections. 

We can also use Eq.~\eqref{eq:limberPk_planes_and_smoothing} to estimate the effect of a finite number of planes.
Results are shown in Fig.~\ref{fig:number_of_planes_th} for different numbers $n$ of planes evenly spaced in comoving distance between the observer and the source. Employing $n=90$ planes yields results that are practically indistinguishable from the limit $n\to\infty$. Even when using only $n=10$ planes, which corresponds to a plane-to-plane distance of $\approx 200\,\Mpc$, the resulting convergence power spectrum deviates by less than $1\,\%$. 

We note however, that this small impact of the number of planes is partly due to our specific choice of source redshift and cosmology. In this case, the integrand in Eq.~\eqref{eq:ps_kappa_1st_order_Limber} is very symmetric for most $\ell$-values of interest, and the integral can be well approximated by a sum of the integrand values at just a few evenly spaced points between the observer and the source. For very different source redshifts (including realistic extended distributions) or cosmologies, the integrand may be more skewed, and the resolution along the line-of-sight may become more important.

\subsubsection{Particle noise and smoothing}
\label{sec:results:systematic_errors:particle_noise_and_smoothing}

\begin{figure}
\centerline{\includegraphics[width=\linewidth]{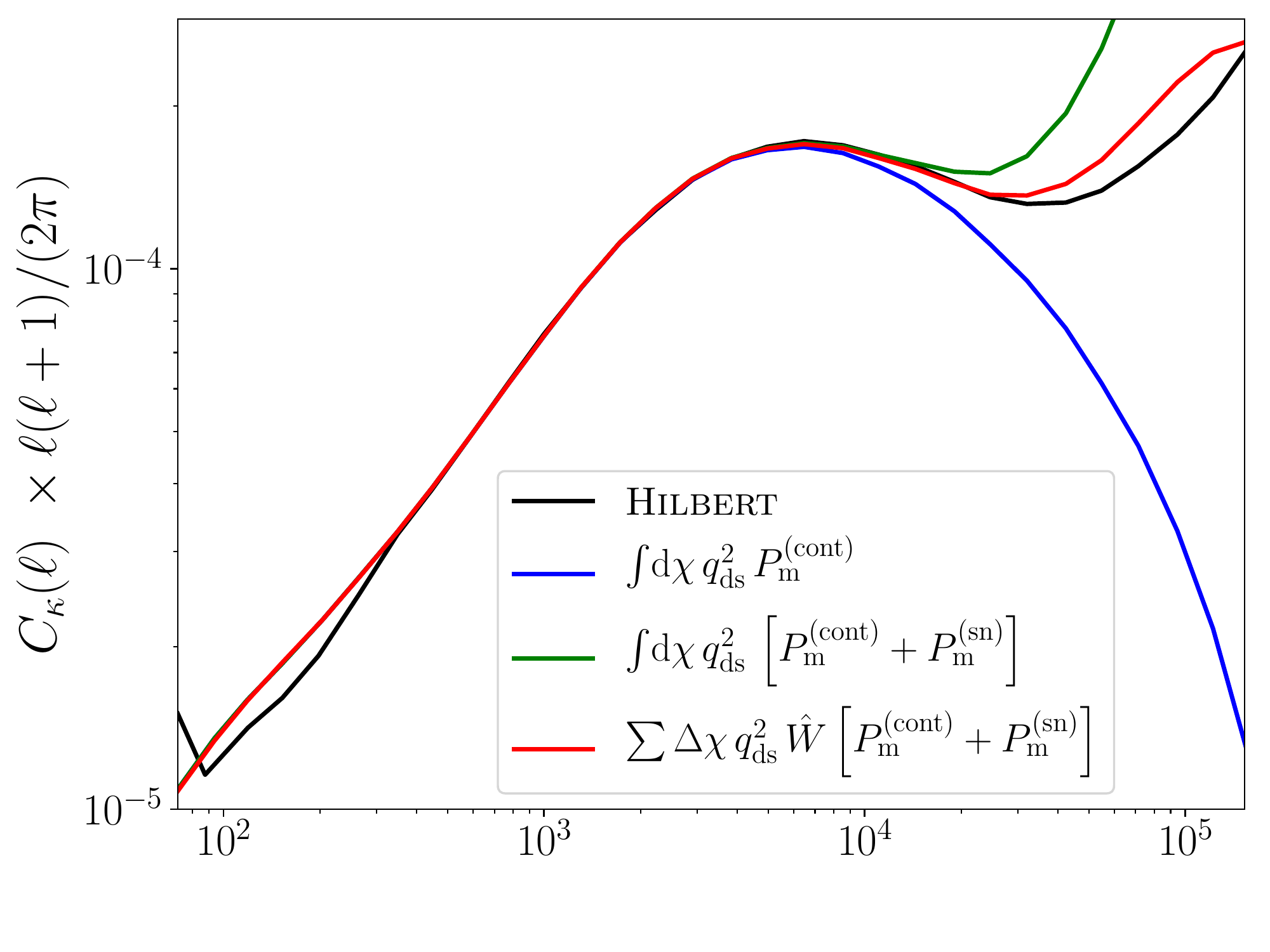}}
\caption{Comparison of convergence power spectra from \hilbert{} (black) with the prediction from Eq.~\eqref{eq:limberPk} only considering the continuous component $P^{\text{(cont)}}_{\text{m}}$ in Eq.~\eqref{eq:P_matter_sim} (blue),  the prediction from Eq.~\eqref{eq:limberPk} also accounting for shot noise (green), as well as the prediction from Eq.~\eqref{eq:limberPk_planes_and_smoothing} taking into account line-of-sight discretization, shot noise, and smoothing (red).}
\label{fig:particle_noise_and_smoothing}
\end{figure}

The three-dimensional matter distribution in the underlying $N$-body simulation lacks power on very large and very small scales due to the finite box size and finite spatial/mass resolution. As mentioned in Sec.~\ref{sec:results:convergence_power_spectra}, this lack of power naturally propagates to the convergence and shear power spectra of the lensing simulations. Moreover, the lensing simulations are affected by additional smoothing, either inherent in some of their processing steps, e.g. when employing meshes of finite resolution, or applied explicitly, e.g. to reduce the impact of shot noise due to finite number of particles in the $N$-body simulation.

The effects of particle shot noise and smoothing are illustrated in Fig.~\ref{fig:particle_noise_and_smoothing} for the \hilbert{} code. Integrating the three-dimensional matter power spectrum directly without accounting for particle shot noise or smoothing yields much smaller values for the convergence power spectrum than the lensing simulation for $\ell \gtrsim 10^4$. Further, accounting for shot noise, but not for smoothing overestimates the convergence power spectra of the lensing simulation.

When the smoothing is taken into account (in addition to the line-of-sight discretization) assuming the kernel of Eq.~\eqref{eq:hilbert_smoothing_kernel}, the prediction based on Eq.~\eqref{eq:limberPk_planes_and_smoothing} agrees with the power spectrum directly measured from the convergence maps to better than $1\%$ for scales $300 \lesssim \ell \lesssim 20000$. This underlines the fact that a good understanding of the amount of shot noise and smoothing carried out by a given lensing code can prove useful in tests of its accuracy.

\subsubsection{Shear power spectra}
\label{sec:results:shear}
\begin{figure}
\centerline{\includegraphics[width=\linewidth]{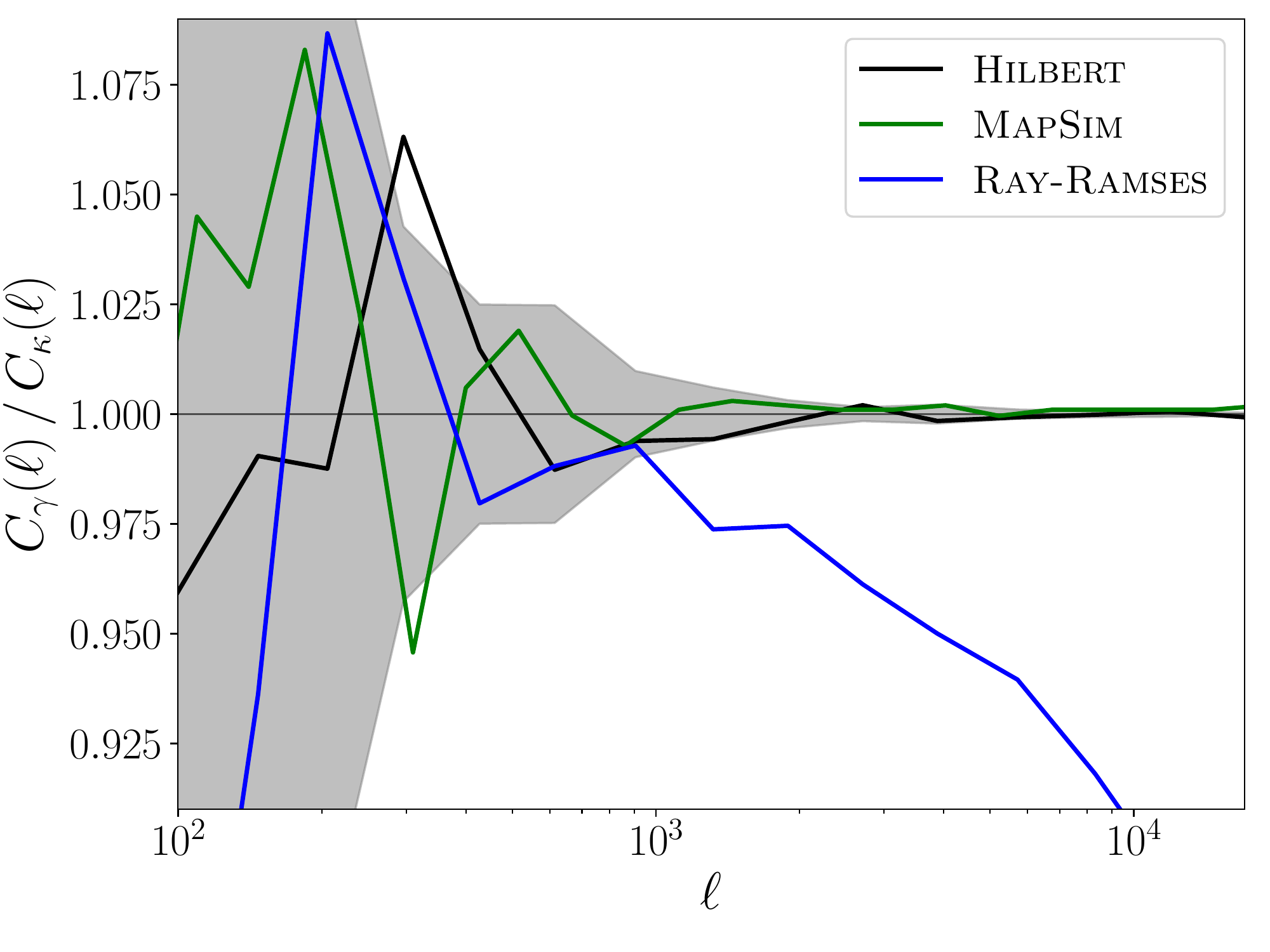}}
\caption{Ratio of shear and convergence power spectra for the and \hilbert{}, \mapsim{}, and \rayramses{} codes (computed from one $10\times10\,\degt^2$ field). The shaded area indicates the expected variance (estimated from the 16 \hilbert{} fields). For $\ell\leq 1000$, the deviations from unity can be attributed to the finite field size. The deviations for $\ell>1000$ for \rayramses{} are due to different effective smoothings for shear and convergence.
}
\label{fig:gamma_to_kappa_spectra}
\end{figure}

We also compare code predictions for the shear power spectrum (not just convergence), since the shear is more directly related to the actually observed galaxy image ellipticities. As mentioned in Section~\ref{sec:theory:wl}, $\harmPS{\text{(EE)}}{\gamma}{\ell}=\harmPS{}{\kappa}{\ell}$ for a flat sky\footnote{On a spherical sky, they differ by a known wave-number-dependent factor that is close to unity for sufficiently large $\ell$.}; this relation holds to high accuracy beyond the Born approximation as well (cf.~Fig.~\ref{fig:born_vs_full_ray_tracing}). In Fig.~\ref{fig:gamma_to_kappa_spectra} we show the ratios of the shear and convergence power spectra for the \hilbert{}, \mapsim{}, and \rayramses{} codes.\footnote{The \hilbert{} and \rayramses{} codes compute the lensing shear directly by projecting the corresponding second-derivatives of the lensing potential. The shear power spectrum from \mapsim{} is calculated from a $\gamma$ map obtained from the $\kappa$ map by Fourier transforming $\tilde{\gamma}(\vell) = (\tilde{\gamma}_1(\vell), \tilde{\gamma}_2(\vell)) = \left(\frac{\ell_1^2 - \ell_2^2}{\ell_1^2 + \ell_2^2}, \frac{2\ell_1\ell_2}{\ell_1^2 + \ell_2^2} \right)\tilde{\kappa}(\vell)$, where $\vell = (\ell_1, \ell_2)$.}
The deviations of the measured ratios from unity seen for $\ell\leq 1000$ can be attributed to the finite size of the field of view.\footnote{For example, different matter structures outside the finite field of view may contribute to the shear seen within the field, but not to the convergence. This can explain why the shear and convergence spectra do not react in the same way when the orientation of the finite field-of-view changes.} On scales $\ell > 1000$, the \rayramses{} result underestimates the shear power spectrum in a non-negligible way: the shear power is smaller by approximately $6\%$ for $\ell \approx 4000$. This difference between the shear and convergence power spectrum can be traced back to an effective larger smoothing that exists in the calculation of the second derivatives of the potential that are integrated to calculate $\gamma_1$ and $\gamma_2$ (cf.~Sec.~\ref{sec:simulations:lensing_simulations:rayramses}). This suppression is therefore expected to remain even if the simulation is performed at higher mass resolution, although the scale at which $\harmPS{\text{(EE)}}{\gamma}{\ell} \neq C_{\kappa}(\ell)$ would be pushed to higher $\ell$ values.

This subtlety in the operation of the \rayramses{} code stresses further the importance to understand the impact of smoothing in lensing simulation codes. The smoothing in \rayramses{} is adaptive, controlled mostly by the resolution of the $N$-body simulation, and as shown here can yield different shear and convergence spectra (even though they should be the same). On the other hand, the smoothing schemes can be partly specified by the user in the other codes (e.g. in the smoothing or mass assignment scheme used to construct the multiple planes/spheres), and should therefore be subject to careful numerical convergence tests.  For the \fabbian{} code, the relation $\harmPS{\text{(EE)}}{\gamma}{\ell} = \harmPS{}{\kappa}{\ell}$ has been verified to hold to sub-percent precision, both in the Born approximation and beyond it \citep{fabbian2018}.

\section{Summary and Conclusions}
\label{sec:conclusions}

In this paper, we have studied the relative accuracy of different weak lensing simulation codes by comparing the statistics of lensing convergence maps that each produced from a common underlying simulation of cosmic large-scale structure. The five codes \hilbert{}, \fabbian{}, \mapsim{}, \mice{} and \rayramses{} we compare (cf.~Table \ref{table:codes} and Sec.~\ref{sec:simulations:lensing_simulations}) were developed independently from each other, and a significant number of results in the literature featuring these codes exists already. The comparison analysis we carried out in this paper thus serves the purpose of cross-checking the validity of these past results, as well as checking the extent to which any systematic difference between codes can affect the analysis of large weak lensing surveys.

A major difference between the codes is how they integrate the rays along the line-of-sight: \hilbert{} and \mapsim{} project the deflector mass field onto planes perpendicular to the central line-of-sight of the field-of-view, \mice{} and \fabbian{} project it onto concentric spherical shells around the observer, and \rayramses{} carries out the lensing integrations using the three-dimensional distribution of the mass without projecting it. Other specifications such as interpolation schemes to reconstruct the deflector mass on regular grids, and additional smoothing applied on these grids, also differ between the codes.

The lensing simulation codes carried out their calculations on the output of the same $N$-body simulation performed with the \ramses{} code. Specifically, the \rayramses{} code ran on-the-fly with the $N$-body calculation and produced the particle snapshots that the remaining codes took as input. We considered a lightcone geometry with area $10\times10\ {\rm deg}^2$, extending out to a single source redshift $\zS = 1$ (cf.~Fig.~\ref{fig:tiling}). This small field of view is not representative of the total area of large surveys like Euclid ($\approx 15000\ {\rm deg}^2$), but is sufficient to compare the code results on small angular scales, which are the scales for which we rely on numerical simulations to resolve nonlinear structure formation.

The main results of the code comparison are:

\begin{itemize}

\item The PDFs of the maps agree to $\approx 1\%$ on $\kappa$ values where the PDF is sizeable, but only after applying  $1\,\arcmint$ smoothing and subtracting the mean convergence over the field of view (cf.~Fig.~\ref{fig:pdfs1}). 

\bigskip

\item The convergence power spectra predicted by the codes agree to $2\%$ for $\ell \lesssim 4000$ (cf.~Fig.~\ref{fig:spectra}). At $\ell = 10^4$, the differences can be as large as $10\%$, mainly due to differences in the smoothing of the matter field.

\bigskip

\item The code predictions for lensing peak counts agree to better than $5\%$ for peaks with signal-to-noise $\nu \lesssim 6$, both for maps with and without shape noise added (cf.~Fig.~\ref{fig:peaks}).

\end{itemize}

Corroborating previous results in the literature, we confirmed the validity of the Born-approximation in the convergence power spectrum from lensing simulations. Following the rays along unperturbed trajectories has an impact that is smaller than $0.2\%$ for $\ell < 10^4$ (see Fig.~\ref{fig:born_vs_full_ray_tracing}). Note, however, that the Born approximation can have a stronger impact on other observables such as galaxy-galaxy lensing, higher order statistics or CMB lensing cross-correlations \citep{2009AaA...499...31H,fabbian2019}.

We also showed that reducing the number of lens planes from 90 (plane thickness $\sim 20\ {\rm Mpc}/h $) to 10 (plane thickness $\sim 200\ {\rm Mpc}/h $) can impact the resulting convergence power spectrum of the \hilbert{} code at the $5\%$ level (see Fig.~\ref{fig:number_of_planes}), albeit with no clear systematic trend. This value is likely exacerbated by the parallel projection this code adopts. Analytic predictions based on matter power spectra suggest that the systematic error due to the line-of-sight discretization is smaller. Moreover, the good agreement between using 90 lens planes or the full $N$-body resolution used by \rayramses{} is telling that the discretization along the line-of-sight is not a critical source of systematic error (at least for slices with width $\lesssim 25\ {\rm Mpc}/h$). Further, we noted that the shear power spectrum predicted by the \rayramses{} code is suppressed relative to that of the convergence on small scales ($\approx 6\%$ on $\ell \approx 4000$, see Fig.~\ref{fig:gamma_to_kappa_spectra}). This is due to an effectively larger smoothing that goes into the calculation of the integrand of $\gamma$ in this code, compared to $\kappa$.

We also compared the convergence power spectrum measured directly from the lensing simulation maps with predictions obtained by analytically integrating the nonlinear three-dimensional matter power spectrum measured from the $N$-body simulation snapshots. Smoothing effects and particle shot noise can be appropriately taken into account analytically. When doing so, e.g., for the \hilbert{} code, the analytical prediction agrees with the lensing simulation spectra to better than $1\%$ out to $\ell = 20000$ (cf.~Fig.~\ref{fig:particle_noise_and_smoothing}). These are scales already well below the scales that our $N$-body simulation could accurately resolve, but an appropriate analytical calculation can capture that loss of resolution and thus be used in self-consistency tests of the codes.

Overall, the comparison of the \hilbert{}, \fabbian{}, \mapsim{}, \mice{} and \rayramses{} codes did not reveal any significant systematic errors in the statistical quantities we analyzed. Further, in comparisons in which larger than a few $\%$ differences were observed, e.g. power spectrum on $\ell \gtrsim 4000$, there are other known sources of larger uncertainty such as the modeling of baryonic processes, or even the accuracy of $N$-body methods in gravity-only simulations. We thus conclude that the current accuracy of weak-lensing simulation codes is acceptable for applications to current and near future data analyses.

In accordance with previous works \citep[e.g.][]{fosalba08}, we find that the convergence and shear power spectra measured from the lensing simulations can be accurately predicted analytically on the relevant scales, which suggests a method to validate lensing codes and simulations. As outlined in Sec.~\ref{sec:theory:wl_sims} and exemplified in Sec.~\ref{sec:results:systematic_errors:particle_noise_and_smoothing}, one adjusts the first-order predictions for the convergence and shear power spectra to account for the peculiarities of the input matter distribution, any line-of-sight discretization, particle noise, and smoothing according to the numerical parameters of cosmic structure simulation and the lensing simulation. One then measures the power spectra from the output of the lensing simulation and compares them to the adjusted first-order predictions. Any significant deviations (e.g. larger than expected from sample variance due to finite area and depth) may then indicate a problem with the lensing simulation code or the quantitative understanding of its numerical properties.

Throughout this paper, we refrained from drawing considerations on the numerical performance of the codes as it is hard to find objective points of comparison. The main distinction in terms of numerical resources concerns the post-processing or on-the-fly nature of the codes. The post-processing codes require relatively low numerical resources in the calculation of the lensing quantities {\it per se}, but the $N$-body data that they analyze (and which is more expensive to generate) is assumed to be pre-existent. Lensing calculations performed on the fly with the $N$-body calculation have the advantage of requiring in principle fewer data storage resources, but are less flexible to changes in the lensing setup adopted (e.g. changes in source redshift may require a new $N$-body simulation). The choices of which method/code will also in general be determined by the specific application in mind, based on the specific features of each code (summarized in Table~\ref{table:codes}). A main conclusion of this work is that, regardless of which method/code is chosen, the accuracy of the result should be within the level of agreement shown in this paper.

As future improvements to the analysis we carried out in this paper, one may extend the comparison to full-sky lensing simulations. Of the codes tested here, only \fabbian{} and \mice{} can currently carry these out. There is in principle no impediment to make \hilbert{}, \mapsim{} and \rayramses{} capable of that too, but that would involve further code development. Additionally, it would be valuable to carry out a similar comparison of numerical codes such as \pinocchio{} \citep{2002MNRAS.331..587M,2017MNRAS.465.4658M}, \peakpatch{} \citep{2019MNRAS.483.2236S}, \icecola{} \citep{2018MNRAS.473.3051I} or that of \cite{giocoli17}, which are capable of generating fast (yet approximate) realizations of the deflector mass distribution and compute their lensing properties. 

\section*{Acknowledgments}

We thank Peter Schneider for useful comments. SH acknowledges support by the DFG cluster of excellence \lq{}Origin and Structure of the Universe\rq{} (\href{http://www.universe-cluster.de}{\texttt{www.universe-cluster.de}}).
AB acknowledges support from the Starting Grant (ERC-2015-STG 678652) "GrInflaGal" of the European Research Council. GF acknowledges support from the CNES postdoctoral fellowship and support by the UK STFC grant ST/P000525/1 as well as by the European Research Council under the European Union's Seventh Framework Programme (FP/2007-2013) ERC Grant Agreement No. [616170].
PF acknowledges support from MINECO through grant ESP2017-89838-C3-1-R, the European Union H2020-COMPET-2017 grant Enabling Weak Lensing Cosmology, and Generalitat de Catalunya through grant 2017-SGR-885.
CG acknowledges the grants ASI n.I/023/12/0, ASI-INAF n.
2018-23-HH.0 and PRIN MIUR 2015 Cosmology and Fundamental Physics: 
illuminating the Dark Universe with Euclid".
MC carried out part of this work while supported by a grant of the EU-ESF, the Autonomous Region of the Aosta Valley and the Italian Ministry of Labour and Social Policy (CUP B36G15002310006), and by a 2019 "Research and Education" grant from Fondazione CRT. The OAVdA is managed by the Fondazione Cl\'{e}ment Fillietroz-ONLUS, which is supported by the Regional Government of the Aosta Valley, the Town Municipality of Nus and the "Unit\'{e} des Communes vald\^{o}taines Mont-\'{E}milius".
CTD is funded by a Science and Technology Facilities Council (STFC) PhD studentship through grant ST/R504725/1.
BL is supported by an ERC Starting Grant, ERC-StG-PUNCA-716532, and is additionally supported by the STFC Consolidated Grants (ST/P000541/1).

The simulations and some of the analyses in this work used the DiRAC facility, managed by the Institute for Computational Cosmology, Durham University on behalf of the STFC DiRAC HPC Facility (\url{www.dirac.ac.uk}). The equipment was funded by BEIS via STFC capital grants ST/K00042X/1, ST/P002293/1, ST/R002371/1 and ST/S002502/1, Durham University and STFC operation grant ST/R000832/1. DiRAC is part of the UK National e-Infrastructure.

\bibliographystyle{mnras}
\bibliography{\bibpath}

\appendix

\bsp  
\label{lastpage}
\end{document}